\documentclass[aps,prd,reprint,groupeaddress,showpacs,preprintnumbers,nofootinbib,nobibnotes,amsmath,amssymb,floatfix,superscriptaddress]{revtex4-1}
\usepackage[colorlinks=true,linkcolor=red,citecolor=blue,urlcolor=blue]{hyperref}
\usepackage{epsfig}
\usepackage{amsmath}
\usepackage{graphicx}
\usepackage{color}
\usepackage{float}
\usepackage{xcolor}
\usepackage{amssymb}
\usepackage{enumitem}
\usepackage{natbib}
\usepackage[T1]{fontenc}

\definecolor{red}{rgb}{0.8,0,0}
\definecolor{violet}{rgb}{0.4,0,0.4}
\definecolor{green}{rgb}{0,0.5,0.0}
\definecolor{navy}{rgb}{0.0,0.0,0.6}
\definecolor{orange}{rgb}{0.8,0.2,0.0}

\usepackage[normalem]{ulem}  

\newcommand{\physrep}{Phys. Rep.}
\newcommand{\apjl}{Astro. Phys. J. Lett.}
\newcommand{\mnras}{Mon. Not. Roy. Astron. Soc.}
\newcommand{\aap}{Astronomy $\&$ Astrophysics}
\newcommand{\nphysa}{Nuc. Phys. A}

\newcommand{\sovast}{Sov. Astron.}

\begin{document}
\title{Massive $\Delta$-resonance admixed hypernuclear stars with anti-kaon condensations}

\author{Vivek Baruah Thapa}
\email{thapa.1@iitj.ac.in}
\author{Monika Sinha}
\email{ms@iitj.ac.in}
\affiliation{Indian Institute of Technology Jodhpur, Jodhpur 342037, India}

\author{Jia Jie Li}
\email{jiajieli@itp.uni-frankfurt.de}
\affiliation{School of Physical Science and Technology, Southwest University,
Chongqing 400700, China}
\affiliation{Institute of Modern Physics, Chinese Academy of Sciences,
Lanzhou 730000, China}
\affiliation{Institute for Theoretical Physics, Goethe University,
Max-von-Laue-Str. 1,
 60438 Frankfurt am Main, Germany}

\author{Armen Sedrakian}
\email{sedrakian@fias.uni-frankfurt.de}
\affiliation{Frankfurt Institute for Advanced Studies, Ruth-Moufang-Stra\ss e, 1, 60438 Frankfurt am Main, Germany}
\affiliation{Institute of Theoretical Physics, University of Wroc\l{}aw, pl. M. Borna 9, 50-204 Wroc\l{}aw, Poland}

\begin{abstract} In this work, we study the effect of (anti)kaon condensation on the properties of compact stars that develop hypernuclear cores with and without an admixture of $\Delta$-resonances.  We work within the covariant density functional theory with the parameters adjusted to $K$-atomic and kaon-nucleon scattering data in the kaonic sector.  The density-dependent parameters in the hyperonic sector are adjusted to the data on $\Lambda$ and $\Xi^-$ hypernuclei data.  The $\Delta$-resonance couplings are tuned to the data obtained from their scattering off nuclei and heavy-ion collision experiments. We find that (anti)kaon condensate leads to a softening of the equation of state and lower maximum masses of compact stars than in the absence of the condensate. Both the $K^-$ and $\bar K^0$-condensations occur through a second-order phase transition, which implies no mixed-phase formation.  For large values of (anti)kaon and $\Delta$-resonance potentials in symmetric nuclear matter, we observe that condensation leads to an extinction of $ \Xi^{-,0}$ hyperons. We also investigate the influence of inclusion of additional hidden-strangeness $\sigma^{*}$ meson in the functional and find that it leads to a substantial softening of the equation of state and delay in the onset of (anti)kaons.  \end{abstract} \keywords{neutron stars; equation of state; (anti)kaon condensates; hyperons; $\Delta$-resonances} \maketitle

\section{Introduction}
\label{sec:intro}

Born in supernova explosions neutron (or compact) stars (NSs) are the densest cosmic bodies in the modern Universe. They provide a unique domain of density range to study the novel states of matter. Indeed, matter in compact stars is compressed by the gravitational force to densities a few times nuclear saturation density, $n_0$~\cite{1996cost.book.....G,Lattimer2016PhR, 2007PrPNP..59...94W,Sedrakian2007PrPNP}.

During the last decade electromagnetic as well as gravitational wave observations placed a number of constraints on the global properties of compact stars (masses, radii, deformabilities, etc.) which significantly constrain the range of admissible equation of state (EoS) models of dense matter. We briefly list below the most important observational results. The masses of massive $M \sim 2 M_{\odot}$ compact star (millisecond pulsars) in binaries with white dwarfs were determined for  J$1614-2230$ ($M=1.908\pm 0.016 M_\odot$) \cite{2010Natur.467.1081D,Arzoumanian_2018}, J$0348+0432$ ($2.01\pm 0.04$ M$_\odot$) \cite{2013Sci...340..448A} and J$0740+6620$ ($2.14^{+0.20}_{-0.18}M_\odot$ with 95$\%$ credibility) \cite{2020NatAs...4...72C}. The radius of a canonical $1.4 M_{\odot}$ compact stars was inferred from low-mass X-ray binaries in globular clusters in the range $10\le R\le 14$~km \cite{2018MNRAS.476..421S}. The mass-radius measurements of PSR J$0030+0451$ by the NICER experiment determined $M=1.44^{+0.15}_{-0.14} M_\odot$, $R=13.02^{+1.24}_{-1.06}$ km \cite{2019ApJ...887L..24M} and $M=1.34^{+0.15}_{-0.16}M_\odot$, $R=12.71^{+1.14}_{-1.19}$ km
(with $68.3\%$ credibility) \cite{2019ApJ...887L..21R}.  The first multimessenger gravitational-wave event GW170817 observed by the LIGO-Virgo collaboration (LVC)~\citep{LIGO_Virgo2017b,LIGO_Virgo2017c,LIGO_Virgo2017a} set constraints on the tidal deformabilities of involved stars which through a tight correlation with the radii predict a radius $12\le R_{1.4}\le 13$~km for the canonical-mass star $M=1.4M_{\odot}$.  The LVC observation of the GW190425 event in gravitational waves determined the component masses range $1.46-1.87 M_{\odot}$ \cite{2020ApJ...892L...3A}. Another event GW190814 suggests a binary with a light component with a mass $2.59^{+0.08}_{-0.09} M_\odot$ \cite{2020ApJ...896L..44A} which falls in the ``mass-gap'' ($2.5 M_\odot\leq M \leq5 M_\odot$). The nature of the lighter companion is still not resolved \cite{2020MNRAS.499L..82M,2020arXiv200706057T,Bombaci2020,Fattoyev2020,Zhang2020}, but the neutron star interpretation appears to be in tension with formation of heavy baryons (hyperons, $\Delta$-resonances) in compact stars~\cite{2020PhRvD.102d1301S,LI2020135812,Dexheimer2020}.

Due to large densities reached in the core region of compact stars, new hadronic degrees of freedom are expected to nucleate in addition to the nucleons. One such possibility is the onset of hyperons, as initially suggested in Ref.~\cite{1960SvA.....4..187A}. This occurs in the inner core of compact stars at about $(2-3)n_0$.  Even though the presence of hyperons in compact stars may seem to be unavoidable, it leads to an incompatibility of the theory with the observations of massive pulsars mentioned above, as is evidenced by many studies
which used either phenomenological \cite{1985ApJ...293..470G, Weber:1989uq, 1995PhRvC..52.3470K, 1997NuPhA.625..435B, 2008PhRvC..78e4306Z} or microscopic \cite{1995PhLB..355...21S, 1998PhRvC..58.3688B, 2000PhRvC..61b5802V, 2009PhRvC..79c4301S,2010PhRvC..81c5803D} approaches.
Specifically, hyperons lead to a softening of the EoS and imply a low value of the maximum mass of compact stars, below those observed. This problem is known as the ``hyperon puzzle''.  The studies prior to the discovery of massive pulsars,
the work during the last decade focused mainly on models which provide sufficient repulsion among the hadronic interactions which guarantees stiffer EoS and larger maximum masses of hypernuclear stars; these have been carried out mostly within the covariant density functional theory~\citep{Weissenborn2012a,Bonanno2012A&A,Colucci_PRC_2013,Dalen2014,Oertel2015, Chatterjee2015,Fortin_PRC_2016,Chen_PRC_2007,Drago_PRC_2014,Cai_PRC_2015, Zhu_PRC_2016,Sahoo_PRC_2018,Kolomeitsev_NPA_2017,Li_PLB_2018,Li2019ApJ, Ribes_2019,Li2020PhRvD}. 
But microscopic models have also been employed~\cite{Yamamoto2016EPJA,Shahrbaf2020PhRvC}.

Another fascinating possibility of the onset of non-nucleonic degrees of freedom is the appearance of stable $\Delta$-resonances in the matter. Whether $\Delta$-resonances play any role in the NSs is still a matter of debate \cite{ Li_PLB_2018,2020PhLB..80235266M}. Early work \cite{1985ApJ...293..470G,1991PhRvL..67.2414G} indicated that the threshold density for the appearance of $\Delta$-resonances could be as high as $(9-10)~n_0$. More recent works \cite{ Li_PLB_2018,2014PhRvD..89d3014D,PhysRevC.90.065809,2015PhRvC..92a5802C} have shown that indeed these non-strange baryons may appear in nuclear matter at density in the range $(1-2)~n_0$. In particular, the recent work which included both hyperons and $\Delta$-resonance \cite{ Li_PLB_2018,Li2019ApJL} showed that the inclusion of $\Delta$-resonances into the NS matter composition reduces the radius of a canonical $1.4 M_{\odot}$ mass compact star, whereas the maximum mass implied by the EoS does not change significantly. The onset of $\Delta$-resonances also shifts the onset of hyperons to higher densities~\cite{2014PhRvD..89d3014D,  Li_PLB_2018,Li2019ApJL}.

Yet another possibility of a new non-nucleonic degree of freedom at high densities is the appearance of various meson (pion, kaon, $\rho$-meson) condensates. Initially, pion-condensation and its implications for neutron star physics was studied~\cite{Migdal_1972,1982ApJ...258..306H,particles2030025}. Later, the focus shifted towards the strangeness-carrying (anti)kaons ($\bar K$) condensate, initially suggested within a chiral perturbative model in Refs.~\cite{1988NuPhA.479..273K,1987PhLB..192..193N}; for further models and developments see~\cite{1994NuPhA.567..937B,1994PhLB..326...14L,1995PhRvC..52.3470K}. It has been then realized that the repulsive optical potential developed by the $K^+$ mesons in the nuclear matter disfavors the presence of kaons in neutron star matter. Several authors \cite{1999PhRvC..60b5803G, 2001PhRvC..63c5802B,1997PhR...280....1P,1996PhRvC..53.1416S,2001PhRvC..64e5805B,Malik:2020jlb,PhysRevD.102.123007} have studied the (anti)kaon condensation in nuclear as well as hypernuclear matter. The onset of (anti)kaons in the compact star matter is very sensitive to the $K^-$ optical potentials as well as the presence of hyperons. In the latter case, it is observed that the threshold density of (anti)kaons is shifted to even higher matter densities \cite{2014PhRvC..90a5801C}.
A generic feature of the onset of the condensates is the softening of the EoS and the reduction of the maximum masses of compact stars, which could become potentially incompatible with the observations of massive pulsars. The onset of (anti)kaon condensation affects many properties of compact stars beyond the equation of state, such as superfluidity~\cite{2018MNRAS.474.3576X}, neutrino emission via direct Urca processes~\cite{2009A&A...506L..13D,2020AdAst2020E..13X}, and bulk viscosity~\cite{Chatterjee_2008}. This is a direct consequence of the changes in the single-particle spectrum of fermions, e.g., the Fermi momenta, effective masses, etc.

In the present work, we explore the possibility of (anti)kaon condensation in $\beta$-equilibrated $\Delta$-resonance admixed hypernuclear matter in the core region of compact stars within the framework of covariant density functional (CDF) model. To construct the EoS, we implement the DD-ME2 parametrization of density functional with density-dependent couplings~\cite{2005PhRvC..71b4312L}. This model has been extended previously to the $\Delta$-resonance admixed hypernuclear matter without (anti)kaon condensation~\cite{ Li_PLB_2018,Li2019ApJL}, showing that the resulting EoS is broadly compatible with the available astrophysical constraints.  It has been also extended to include the effect of strong magnetic fields~\cite{particles3040043}. This work, therefore, will focus on the novel aspects that are introduced by the (anti)kaon condensation.

The paper is arranged as follows. In Sec.~\ref{sec:formalism} we briefly describe the density-dependent CDF formalism and its extension to (anti)kaons condensation in $\Delta$-resonances admixed hypernuclear matter. Sec.~\ref{sec:results} is devoted to numerical results and their discussions. The conclusions and future perspectives are given in Sec.~\ref{sec:conclusions}. We use natural units $\hbar=c=1$ throughout.

\section{Formalism} \label{sec:formalism}

\subsection{Density Dependent CDF Model}

In this work, we consider the density dependent CDF model to study the transition of matter from hadronic to (anti)kaon condensed phase in $\beta$-equilibrated $\Delta$-resonance admixed hypernuclear matter. The matter composition is considered to be of the baryon octet ($b\equiv N,\Lambda,\Sigma,\Xi$), $\Delta$-resonances ($d\equiv \Delta^{++},\Delta^+,\Delta^0,\Delta^-$), (anti)kaons ($\bar{K}\equiv K^-, \bar{K}^0$) alongside leptons ($l$) such as electrons and muons. The strong interactions between the baryons as well as the (anti)kaons are mediated by the isoscalar-scalar $\sigma$, $\sigma^*$, isoscalar-vector $\omega^\mu$, $\phi^{\mu}$ and isovector-vector $\boldsymbol{\rho}^{\mu}$ meson fields. The additional hidden strangeness mesons ($\sigma^*, \phi^\mu$) are considered to mediate the hyperon-hyperon as well as (anti)kaon-hyperon interactions. The total Lagrangian density consisting of the baryonic, leptonic and kaonic parts is given by \cite{1999PhRvC..60b5803G,2000NuPhA.674..553P,2001PhRvC..63c5802B,Li_PLB_2018}
\begin{equation} \label{eqn.1} 
\begin{aligned}
\mathcal{L} & = \sum_{b} \bar{\psi}_b(i\gamma_{\mu} D^{\mu}_{(b)} - m^{*}_b) \psi_b + \sum_{d} \bar{\psi}_{d\nu}(i\gamma_{\mu} D^{\mu}_{(d)} - m^{*}_d) \psi^{\nu}_{d} \\ & + \sum_{l} \bar{\psi}_l (i\gamma_{\mu} \partial^{\mu} - m_l)\psi_l + D^{(\bar{K})*}_\mu \bar{K} D^\mu_{(\bar{K})} K - m^{*^2}_K \bar{K} K \\
 & + \frac{1}{2}(\partial_{\mu}\sigma\partial^{\mu}\sigma - m_{\sigma}^2 \sigma^2) + \frac{1}{2}(\partial_{\mu}\sigma^*\partial^{\mu}\sigma^* - m_{\sigma^*}^2 \sigma^{*2})  \\
 & -  \frac{1}{4}\omega_{\mu\nu}\omega^{\mu\nu} + \frac{1}{2}m_{\omega}^2\omega_{\mu}\omega^{\mu} - \frac{1}{4}\boldsymbol{\rho}_{\mu\nu} \cdot \boldsymbol{\rho}^{\mu\nu} + \frac{1}{2}m_{\rho}^2\boldsymbol{\rho}_{\mu} \cdot \boldsymbol{\rho}^{\mu} \\
& - \frac{1}{4}\phi_{\mu\nu}\phi^{\mu\nu} + \frac{1}{2}m_{\phi}^2\phi_{\mu}\phi^{\mu}
\end{aligned} \end{equation}
where the fields $\psi_b$, $\psi^{\nu}_d$, $\psi_l$ correspond to the baryon octet, $\Delta$-baryon and lepton fields. $m_b$, $m_d$, $m_K$ and $m_l$ represent the bare masses of members of baryon octet, $\Delta$-quartet, isospin doublet for (anti)kaons and leptons respectively. The covariant derivative in Eq.\eqref{eqn.1} is
\begin{equation}\label{eqn.2}
D_{\mu (j)} = \partial_\mu + ig_{\omega j} \omega_\mu + ig_{\rho j} \boldsymbol{\tau}_j \cdot \boldsymbol{\rho}_{\mu} + ig_{\phi j} \phi_\mu
\end{equation}
with $j$ denoting the baryons ($b,d$) and (anti)kaons ($\bar{K}$). The density-dependent coupling constants are denoted by $g_{pj}$ where `$p$' index labels the mesons.
The isospin operator for the isovector-vector meson fields is represented by $\tau_j$.
The gauge field strength tensors for the vector meson fields are given by
 \begin{equation} \label{eqn.3}
  \begin{aligned}
  \omega_{\mu \nu} & = \partial_{\nu}\omega_{\mu} - \partial_{\mu}\omega_{\nu} ,\\
  \boldsymbol{\rho}_{\mu \nu} & = \partial_{\nu}
  \boldsymbol{\rho}_{\mu} - \partial_{\mu}\boldsymbol{\rho}_{\nu}, \\
  \phi_{\mu \nu} & = \partial_{\nu}\phi_{\mu} - \partial_{\mu}\phi_{\nu} .
\end{aligned}
\end{equation}

The Dirac effective baryon and (anti)kaon masses in Eq.\eqref{eqn.1} are given by
\begin{equation} \label{eqn.4}
\begin{aligned}
    m_{b}^* & = m_b - g_{\sigma b}\sigma - g_{\sigma^* b}\sigma^*,\quad
    m_{d}^* = m_d - g_{\sigma d}\sigma, \\
    m_{K}^* & = m_K - g_{\sigma K}\sigma - g_{\sigma^* K}\sigma^*
  \end{aligned}
\end{equation}
In the relativistic mean-field approximation, the meson fields obtain
expectation values which are  given by 
\begin{widetext}
\begin{equation} \label{eqn.5}
\begin{aligned}
\sigma & = \sum_{b} \frac{1}{m_{\sigma}^2} g_{\sigma b}n_{b}^s + \sum_{d} \frac{1}{m_{\sigma}^2} g_{\sigma d}n_{d}^s + \sum_{\bar{K}} \frac{1}{m_{\sigma}^2} g_{\sigma K}n_{\bar{K}}^s, \quad
\sigma^* = \sum_{b} \frac{1}{m_{\sigma^*}^2} g_{\sigma^* b}n_{b}^s + \sum_{\bar{K}} \frac{1}{m_{\sigma^*}^2} g_{\sigma^* K}n_{\bar{K}}^s,\\
  \omega_{0} & = \sum_{b} \frac{1}{m_{\omega}^2} g_{\omega b}n_{b} + \sum_{d} \frac{1}{m_{\omega}^2} g_{\omega d}n_{d} - \sum_{\bar{K}} \frac{1}{m_{\omega}^2} g_{\omega K}n_{\bar{K}}, \quad
  \phi_{0} = \sum_{b} \frac{1}{m_{\phi}^2} g_{\phi b}n_{b} - \sum_{\bar{K}} \frac{1}{m_{\phi}^2} g_{\phi K}n_{\bar{K}} \\
    \rho_{03} & = \sum_{b} \frac{1}{m_{\rho}^2} g_{\rho b}
  \boldsymbol{\tau}_{b3}n_{b} + \sum_{d} \frac{1}{m_{\rho}^2} g_{\rho d}
  \boldsymbol{\tau}_{d3}n_{d}
   + \sum_{\bar{K}} \frac{1}{m_{\rho}^2} g_{\rho K} \boldsymbol{\tau}_{\bar{K}3}n_{\bar{K}}
\end{aligned}
\end{equation}
\end{widetext}
where $n^s= \langle\bar{\psi} \psi \rangle$ and $n=\langle\bar{\psi} \gamma^0 \psi \rangle$ denote the scalar and vector (number) densities respectively. The explicit form of scalar and vector density of baryons in the $T=0$ limit is
\begin{equation} \label{eqn.6}
\begin{aligned}
n^{s}_j & = \frac{2J_j + 1}{4 \pi^2} m^{*}_{j} \left[ p_{{F}_{j}} E_{F_j}  - m_{j}^{*^2} \ln \left( \frac{p_{{F}_j} + E_{F_j}}{m_{j}^{*}} \right) \right], \\
n_j & = \frac{2J_j + 1}{6 \pi^2}p_{{F}_{j}}^{3}
\end{aligned}
\end{equation}
respectively with $J_j$, $p_{{F}_{j}}$ and $E_{F_j}$ being the spin, Fermi momentum and Fermi energy of the $j$-th baryon. For the case of $s$-wave (anti)kaons, the number density is given as
\begin{equation} \label{eqn.7}
\begin{aligned}
    n_{K^-, \bar{K}^0} & = 2 \left( \omega_{\bar{K}} + g_{\omega K} \omega_0 + g_{\phi K} \phi_0 \pm \frac{1}{2} g_{\rho K} \rho_{03} \right) \\
    & = 2 m^*_K \bar{K} K.
\end{aligned}
\end{equation}
Here, $\omega_{\bar{K}}$ represents the in-medium energies of (anti)kaons and are given by (considering isospin projection as $\mp 1/2$ for $K^-, \bar{K}^0$)
\begin{equation} \label{eqn.8}
    \omega_{K^{-} , \bar{K}^0} = m^*_K - g_{\omega K} - g_{\phi K} \phi_0 \mp \frac{1}{2} g_{\rho K} \rho_{03}.
\end{equation}
In case of leptons ($l$), the number density is given by $n_l = p_{{F}_{l}}^{3}/3 \pi^2$. The chemical potential of the $j$-th baryon is 
\begin{equation}\label{eqn.9}
\begin{aligned}
	& \mu_{j} = \sqrt{p_{F_j}^2 + m_{j}^{*2}} + \Sigma_{B},
\end{aligned}
\end{equation}
where $\Sigma_B = \Sigma^{0} + \Sigma^{r}$ denotes the vector self-energy with
\begin{equation}\label{eqn.10}
\begin{aligned}
	\Sigma^{0} & = g_{\omega j}\omega_{0} + g_{\phi j}\phi_{0} + g_{\rho j} \boldsymbol{\tau}_{j3} \rho_{03},
\end{aligned}
\end{equation}
\begin{equation}
\begin{aligned}\label{eqn.rear}
	\Sigma^{r} & = \sum_{b} \left[ \frac{\partial g_{\omega b}}{\partial n}\omega_{0}n_{b} - \frac{\partial g_{\sigma b}}{\partial n} \sigma n_{b}^s + \frac{\partial g_{\rho b}}{\partial n} \rho_{03} \boldsymbol{\tau}_{b3} n_{b} \right. \\
	& \left. + \frac{\partial g_{\phi b}}{\partial n}\phi_{0}n_{b} \right] + \sum_{d} (\psi_b \longrightarrow \psi_{d}^{\nu}).
\end{aligned}
\end{equation}
Eq.\eqref{eqn.rear} is the rearrangement term which is required in case of density-dependent meson-baryon coupling models to maintain the thermodynamic consistency \cite{2001PhRvC..64b5804H}. Here, $n=\sum_{j} n_j$ represents the total baryon number density.

The threshold condition for the onset of $j$-th baryon into the nuclear matter is given by \cite{2001PhRvC..63c5802B}
\begin{equation} \label{eqn.11}
\mu_j = \mu_n - q_j \mu_e
\end{equation}
where $q_j$ is the charge of the $j$-th baryon. $\mu_e= \mu_n -\mu_p$ is the chemical potential of electron with $\mu_n$, $\mu_p$ denoting the same for neutron and proton. With increasing density, the Fermi energy of electrons increases and once it reaches the rest mass of muons i.e. $\mu_e = m_{\mu}$, muons start to appear in the nuclear matter.

In case of (anti)kaons, the threshold conditions are governed by the strangeness changing processes such as, $N \rightleftharpoons N + \bar{K}$ and $e^- \rightleftharpoons K^-$ \cite{1997PhR...280....1P,1996cost.book.....G} and are given by
\begin{equation} \label{eqn.12}
\begin{aligned}
    \mu_n - \mu_p = \omega_{K^-} = \mu_e, \quad \omega_{\bar{K}^0} = 0.
\end{aligned}
\end{equation}

The total energy density due to the fermionic part is given by
\begin{widetext}
\begin{eqnarray} \label{eqn.13}
\begin{aligned}
\varepsilon_f & = \frac{1}{2}m_{\sigma}^2 \sigma^{2} + \frac{1}{2} m_{\omega}^2 \omega_{0}^2 + \frac{1}{2}m_{\rho}^2 \rho_{03}^2 + \sum_{j\equiv b,d} \frac{2J_j + 1}{2 \pi^2} \left[ p_{{F}_j} E^3_{F_j} - \frac{m_{j}^{*2}}{8} \left( p_{{F}_j} E_{F_j} + m_{j}^{*2} \ln \left( \frac{p_{{F}_j} + E_{F_j}}{m_{j}^{*}} \right) \right) \right] \\
	 & + \frac{1}{2}m_{\sigma^*}^2 \sigma^{*2} + \frac{1}{2} m_{\phi}^2 \phi_{0}^2 + \frac{1}{\pi^2}\sum_l \left[ p_{{F}_l} E^3_{F_l} - \frac{m_{l}^{2}}{8} \left( p_{{F}_l} E_{F_l} + m_{l}^{2} \ln \left( \frac{p_{{F}_l} + E_{F_l}}{m_{l}} \right) \right) \right].
\end{aligned}
\end{eqnarray}
\end{widetext}
And the energy density contribution from the kaonic matter is
\begin{equation} \label{eqn.14}
    \varepsilon_{\bar{K}} = m^*_K (n_{K^-} + n_{\bar{K}^0})
  \end{equation}
giving the total energy density as $\varepsilon = \varepsilon_{\bar{K}}+\varepsilon_f$. Now, because (anti)kaons being bosons are in the condensed phase at $T=0$, the matter pressure is provided only by the baryons and leptons and is given by the Gibbs-Duhem relation
\begin{equation}\label{eqn.15}
    p_m = \sum_{j\equiv b,d} \mu_j n_j + \sum_{l} \mu_l n_l - \varepsilon_f.
\end{equation}
The rearrangement term in Eq.~\eqref{eqn.rear} contributes explicitly to the matter pressure term only through the vector self-energy term.

Two additional constraints --  the charge neutrality and global baryon number conservation -- should be taken into account to calculate the equation of state self-consistently. 
The charge neutrality condition is given by
\begin{eqnarray} \label{eqn.16}
    \sum_b q_b n_b + \sum_d q_d n_d - n_{K^-} - n_e - n_\mu = 0.
\end{eqnarray}

\subsection{Coupling parameters}

In the density dependent CDF model implemented in this work, DD-ME2 \cite{2005PhRvC..71b4312L} coupling parametrization is incorporated. The coupling functional dependence of the scalar $\sigma$ and vector $\omega$-meson on density is given by
\begin{equation}\label{eqn.17}
g_{i N}(n)= g_{i N}(n_{0}) f_i(x), \quad \quad \text{for }i=\sigma,\omega,
\end{equation}
where, $x=n/n_0$, $n$, $n_0$ being the total baryon number density and nuclear saturation density respectively with
\begin{equation}\label{eqn.18}
f_i(x)= a_i \frac{1+b_i (x+d_i)^2}{1+c_i (x+d_i)^2}
\end{equation}

For the case with $\rho$-meson, the density dependence coupling functional is defined as
\begin{equation}\label{eqn.19}
g_{\rho N}(n)= g_{\rho N}(n_{0}) e^{-a_{\rho}(x-1)}
\end{equation}
The parameters of the meson-nucleon couplings in DD-ME2
parametrization model is given in Table \ref{tab:1}. The coefficients
associated with DD-ME2 model are fitted to reproduce nuclear
phenomenology; the details of which can be found in
Ref.~\cite{2005PhRvC..71b4312L}.  
Since the nucleons do not couple to the strange mesons, $g_{\sigma^*
  N}=g_{\phi N}=0$.
\begin{table} [h!]
\centering
\caption{The meson masses and parameters of the DD-ME2 parametrization used in Eq.~\eqref{eqn.17}
  and \eqref{eqn.18}.}
\begin{tabular}{ccccccc}
\hline \hline
Meson ($i$) & $m_i$ (MeV) & $a_{i}$ & $b_{i}$ & $c_{i}$ & $d_{i}$ & $g_{iN}$ \\
\hline
$\sigma$ & 550.1238 & 1.3881 & 1.0943 & 1.7057 & 0.4421 & 10.5396 \\
$\omega$ & 783 & 1.3892 & 0.9240 & 1.4620 & 0.4775 & 13.0189 \\
$\rho$ & 763 & 0.5647 & & & & 7.3672 \\
\hline
\end{tabular}
\label{tab:1}
\end{table}
The masses of the additional hidden strangeness mesons are taken as $m_{\sigma^*}=975$ MeV and $m_{\phi}=1019.45$ MeV. The nuclear saturation properties are provided in Table~\ref{tab:2}. The parameters $E_0$, $K_0$, $Q_0$ denote the saturation energy, incompressibility, and skewness in isoscalar sector, and    $E_{sym}$, $L_{sym}$ for symmetry energy coefficient and its slope in isoscalar sector, all evaluated at the saturation density. 
It should be noted, that the
  experimentally obtained values of some of these parameters have 
  an uncertainty range given by $n_0 \in [0.14-0.17]$ fm$^{-3}$~\cite{2018PhRvC..97b5805M}, $-E_0\in [15-17]$ MeV~\cite{2018PhRvC..97b5805M}, $K_0\in [220-260]$ MeV~\cite{2006EPJA...30...23S,2010JPhG...37f4038P}, $E_{sym}\in [28.5-34.9]$
  MeV~\cite{2013PhLB..727..276L,2017RvMP...89a5007O}. Once the parameters of the
  model are fixed to particular values within the range indicated above, the EoS is obtained by straightforward extrapolation to the high-density regime. At present, the high-density properties of dense matter are constrained
  by astrophysics of compact stars and modeling of heavy-ion collision experiments,
  both of which carry uncertainties of their own.

\begin{table} [h!]
\centering
\caption{The nuclear properties of the density-dependent CDF model (DD-ME2) at $n_0$.}
\begin{tabular}{ccccccc}
\hline \hline
 $n_0$ & $E_0$ & $K_0$ & $Q_0$ & $E_{sym}$ & $L_{sym}$ & $m^*_N/m_N$ \\
 (fm$^{-3}$) & (MeV) & (MeV) & (MeV) & (MeV) & (MeV) & \\
\hline
 0.152 & $-16.14$ & 250.9 & 478.9 & 32.3 & 51.3 & 0.57 \\
\hline
\end{tabular}
\label{tab:2}
\end{table}
The bare masses of the members of the baryon octet, $\Delta$-quartet and (anti)-kaons considered in this work are, $m_{\Lambda}=1115.68$ MeV, $m_{\Xi^0}=1314.86$ MeV, $m_{\Xi^-}=1321.71$ MeV, $m_{\Sigma^+}=1189.37$ MeV, $m_{\Sigma^0}=1192.64$ MeV, $m_{\Sigma^+}=1197.45$ MeV, $m_{\Delta}=1232$ MeV, $m_{K}=493.69$ MeV.

For the meson-hyperon vector coupling parameters, we incorporated the SU(6) symmetry and quark counting rule \cite{2001PhRvC..64e5805B,2011PhRvC..84c5809R} as
\begin{equation}\label{eqn.20}
\begin{aligned}
\frac{1}{2}g_{\omega \Lambda} & = \frac{1}{2}g_{\omega \Sigma}=g_{\omega \Xi}= \frac{1}{3}g_{\omega N}, \\
2g_{\phi \Lambda} & =2g_{\phi \Sigma}=g_{\phi \Xi}= -\frac{2\sqrt{2}}{3}g_{\omega N}, \\
\frac{1}{2}g_{\rho \Sigma} & =g_{\rho \Xi}=g_{\rho N}, \ \ g_{\rho \Lambda}=0.
\end{aligned}
\end{equation}
The scalar meson-hyperon couplings are calculated by considering the optical potentials of $\Lambda$, $\Sigma$, $\Xi$ as $-30$ MeV, $+30$ MeV and $-14$ MeV respectively \cite{particles3040043}. Furthermore, the scalar strange meson $\sigma^*$-hyperon coupling is evaluated from the measurements on light double-$\Lambda$ nuclei and fitted to the optical potential depth $U^{\Lambda}_{\Lambda}(n_0/5)=-0.67$ MeV \cite{2018EPJA...54..133L}. The scalar meson-hyperon couplings for the other strange baryons can be obtained from the relationship,
\begin{equation}\label{eqn.21}
\frac{g_{\sigma^* Y}}{g_{\phi Y}}=\frac{g_{\sigma^* \Lambda}}{g_{\phi
    \Lambda}},
\ \ Y \in \text{$\{\Xi,\Sigma\}$}.
\end{equation}
Table~\ref{tab:3} provides the numerical values of the meson-hyperon couplings at nuclear saturation density,
where $R_{\sigma Y}=g_{\sigma Y}/g_{\sigma N}$, $R_{\sigma^* Y}=g_{\sigma^* Y}/g_{\sigma N}$ denote the scaling factors for non-strange and strange scalar mesons coupling to hyperons respectively.
\begin{table}[h!]
\centering
\caption{Scalar meson-hyperon coupling constants for DD-ME2 parametrization.}
\begin{tabular}{cccc}
\toprule
 & $\Lambda$ & $\Xi$ & $\Sigma$ \\
\hline
$R_{\sigma Y}$ & 0.6105 & 0.3024 & 0.4426 \\
$R_{\sigma^* Y}$ & 0.4777 & 0.9554 & 0.4777 \\
\hline
\end{tabular}
\label{tab:3}
\end{table}
Because experimental information on the $\Delta$-resonance is scarce, the meson-$\Delta$ baryon couplings are treated as parameters. In the subsequent discussion we consider $g_{\omega d}=1.10~g_{\omega N}$ and $g_{\rho d}= g_{\rho N}$ for vector-meson couplings \cite{ Li_PLB_2018,2019PhRvC.100a5809L}.  For the scalar meson-$\Delta$ baryon couplings we use two values of the isoscalar potentials viz. $V_{\Delta}= V_{N}$ and $5/3~V_{N}$ with $V_N$ being the nucleon potential \cite{ Li_PLB_2018,LI2020135812}.  These values were extracted from the studies of electron and pion scattering off nuclei studies as well as studies of heavy-ion-collisions which involved $\Delta$-resonance production.

The numerical values of scalar meson-$\Delta$-baryon coupling parameters with $V_{\Delta}= V_N$ is $R_{\sigma d}=1.10$ and that with $V_{\Delta}= 5/3~V_N$ is $R_{\sigma d}=1.23$, where  $R_{\sigma d}= g_{\sigma d}/g_{\sigma N}$ denotes the non-strange scalar meson coupling to $\Delta$-resonances. Similar to the nucleons, $\Delta$-resonances do not couple to $\sigma^*$, $\phi$-mesons, i.e, $g_{\sigma^* d}= g_{\phi d}=0$.

The meson-(anti)kaon couplings are fixed according to
Refs.~\cite{2013PhRvC..87d5802G,2014PhRvC..90a5801C} and are taken as
density indepedent.  The vector meson-(anti)kaon coupling parameters
are evaluated from the isospin counting rule
\cite{2014PhRvC..90a5801C} and are given as
\begin{equation}\label{eqn.22}
g_{\omega K} = \frac{1}{3} g_{\omega N}, \quad g_{\rho K} = g_{\rho N}.
\end{equation}
And in case of the additional hidden strange force mediating mesons, the couplings are given as \cite{2001PhRvC..64e5805B}
\begin{equation}\label{eqn.23}
g_{\sigma^* K} = 2.65, \quad g_{\phi K} = 4.27.
\end{equation}

The scalar meson-(anti)kaon coupling constants are calculated by fitting to the real part of $K^-$ optical potential at nuclear saturation density. The readers may refer to Ref.~\cite{PhysRevD.102.123007} for details.
Refs.~\cite{1997NuPhA.625..287W,1999PhRvC..60b4314F, particles2030025} show that the (anti)kaons experience an attractive potential in nuclear matter whereas the opposite is true for the case with kaons in nuclear matter \cite{1997NuPhA.625..372L,2000PhRvC..62f1903P}. Different model calculations \cite{1994PhLB..337....7K, 1997NuPhA.625..287W,1998PhLB..426...12L,2000NuPhA.669..153S,1999PhRvC..60b4314F} have provided the $K^-$ optical potential in normal nuclear matter to be in the range from $-40$ MeV to $-200$ MeV. We have chosen a $K^-$ optical potential range of $-120\leq U_{\bar{K}}\leq -150$ MeV in this work and numerical
values of $g_{\sigma K}$  for the mentioned optical potential range is provided in Table~\ref{tab:4}.
\begin{table} [h!]
\centering
\caption{Scalar $\sigma$ meson-(anti)kaon coupling parameter values in DD-ME2 parametrization at $n_0$.}
\begin{tabular}{ccccc}
\hline \hline
$U_{\bar{K}}$ (MeV) & $-120$ & $-130$ & $-140$ & $-150$ \\
\hline
$g_{\sigma K}$ & 0.4311 & 0.6932 & 0.9553 & 1.2175 \\
\hline
\end{tabular}
\label{tab:4}
\end{table}

\section{Results} \label{sec:results}

In this section we report our numerical results for matter composition with (anti)kaons and (a) Nucleons + Hyperons (NY), (b) Nucleons + Hyperons + $\Delta$-resonances (NY$\Delta$) for varying values of (anti)kaon optical potentials.  The case of pure nuclear matter with (anti)kaons was considered already in Ref.~\cite{PhysRevD.102.123007} and the reader is referred to that work.  From calculations, it is found that the phase transition to (anti)kaon condensed phase is of the second-order for both NY and NY$\Delta$ compositions. In all the calculations the $K^-$-meson is observed to appear before the onset of $\bar{K}^0$. Table~\ref{tab:6} provides the threshold densities of (anti)kaons for different values of $\Delta$-baryon as well as $U_{\bar{K}}$ potentials for two matter compositions.  \begin{figure} [h!]
  \begin{center}
\includegraphics[width=8.5cm,keepaspectratio]{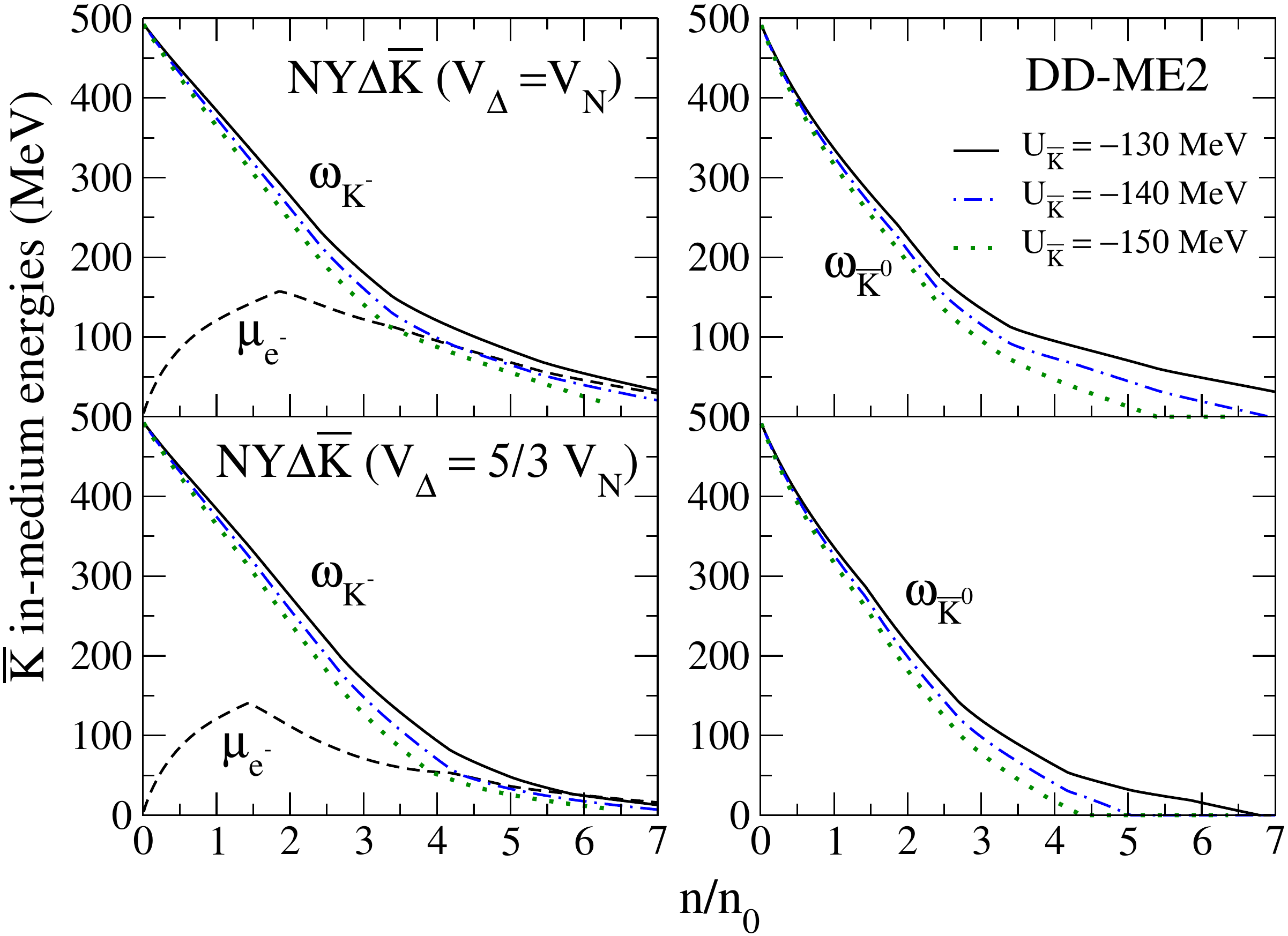}
\caption{The effective energy of (anti)kaons as a function of baryon number density in NY$\Delta$ matter for $\Delta$-potential values $V_{\Delta}=1$ (top panels) and $5/3~V_N$ (bottom panels). Left and right panels
  show the energies of  $K^-$ and $\bar{K}^0$ respectively. The chemical potential of electron for the
  same matter composition is depicted by the dashed curve. The solid, dash-dotted, dotted lines represent the $U_{\bar{K}}$ values of $-130,-140,-150$ MeV respectively.}
\label{fig-4}
\end{center}
\end{figure}

\begin{figure*} [t!]
  \begin{center}
\includegraphics[width=14.5cm,keepaspectratio ]{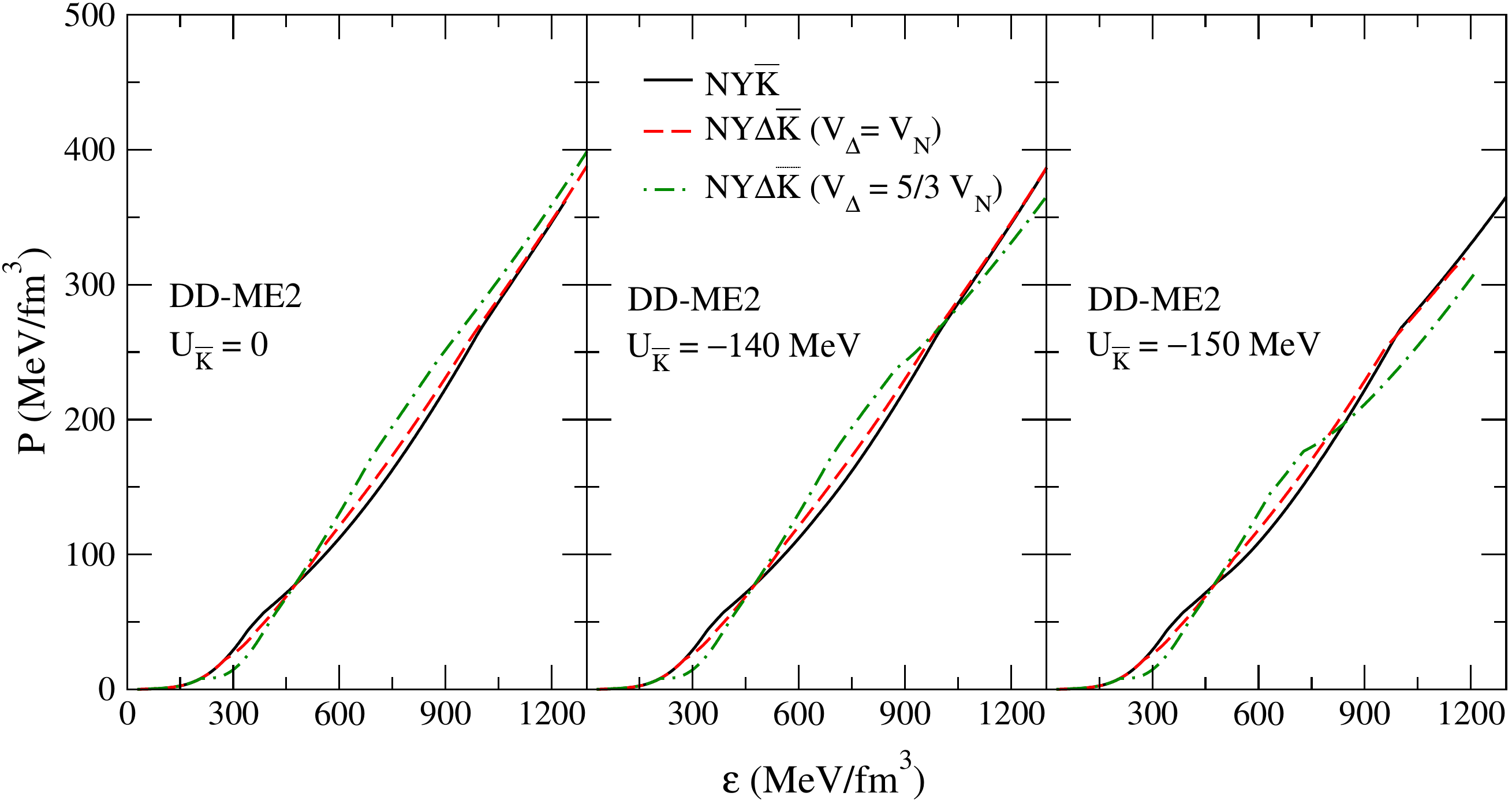}
\caption{Pressure as a function of energy  density (EoS) for zero-temperature, charge-neutral   NY matter (solid lines), NY$\Delta$ matter
  with $\Delta$-potential $V_\Delta=V_N$ (dashed lines) and $V_\Delta=5/3~V_N$  (dash-dotted lines). The three panels correspond to different values of (anti)kaon potential: $U_{\bar{K}}=0$ (left panel), $U_{\bar{K}}=-140$
  (middle panel), and $U_{\bar{K}}=-150$ MeV (right panel).  }
\label{fig-1}
\end{center}
\end{figure*}
It is observed that the (anti)kaons do not appear at all in case of $U_{\bar{K}}=-120$ MeV for all matter compositions. (Anti)kaons are observed to appear only after $U_{\bar{K}}=-130$ MeV with $V_{\Delta}=5/3~V_N$. This happens as the higher $\Delta$-potential shifts the onset of hyperons to higher densities making the way for the (anti)kaons. In all the cases considered, it is observed that with the inclusion of $\Delta$-resonances into the composition of matter the threshold densities of onset of (anti)kaon is shifted to higher densities.

\begin{table} [h!]
\centering
\caption{Threshold densities, $n_{u}$ for (anti)kaon condensation in NY and NY$\Delta$
  matter for different values of $\Delta$-potentials and $K^-$ optical potential depths $U_{\bar{K}}(n_0)$.}
\begin{tabular}{c|cc|cc|cc}
\hline \hline
Config. & \multicolumn{2}{c}{NY$\bar{K}$} & \multicolumn{4}{|c}{NY$\Delta \bar{K}$} \\
 \cline{1-7}
 & & & \multicolumn{2}{c|}{$V_{\Delta}=V_N$} & \multicolumn{2}{c}{$V_{\Delta}= 5/3~V_N$} \\
 \cline{4-7}
$U_{\bar{K}}$ & $n_{u}$($K^-$) & $n_{u}$($\bar{K}^0$) & $n_{u}$($K^-$) & $n_{u}$($\bar{K}^0$) & $n_{u}$($K^-$) & $n_{u}$($\bar{K}^0$) \\
(MeV) & ($n_0$) & ($n_0$) & ($n_0$) & ($n_0$) & ($n_0$) & ($n_0$) \\
\hline
$-120$ & $-$ & $-$ & $-$ & $-$ & $-$ & $-$ \\
$-130$ & $-$ & $-$ & $-$ & $-$ & 5.86 & 6.79 \\
$-140$ & 3.97 & 6.95 & 4.26 & 6.92 & 4.37 & 5.05 \\
$-150$ & 3.06 & 5.59 & 3.33 & 5.39 & 3.90 & 4.37 \\
\hline
\end{tabular}
\label{tab:6}
\end{table}

Figure~\ref{fig-4} shows the in-medium (effective) energies of $\bar{K}$ mesons as a function of baryon (vector) number density in NY$\Delta $ matter described by the DD-ME2 CDF. The onset of $K^-$ mesons condense in the compact star matter occurs when the respective effective energy crosses the electron chemical potential, which then marks the threshold density. In the case $\bar{K}^0$ mesons, the condensate appears when their in-medium energy value reaches zero. With the increase in the values of $U_{\bar{K}}$, the density threshold for the onset of the (anti)kaons is shifted to lower densities.

The EoSs with NY and NY$\Delta$ matter compositions in the absence as well as in presence of (anti)kaon degrees of freedom are shown in Fig.~\ref{fig-1}. In the case with no (anti)kaons in the matter, the EoSs of NY$\Delta$ matter is  stiffer than the EoS of  NY matter in the high-density regime and the opposite is true in the low-density regime.
This is consistent with the results of Ref.~\cite{Li_PLB_2018} found within the same DD-ME2 parametrization. 

The middle and right panels of Fig.~\ref{fig-1} include (anti)kaons with potential values $U_{\bar{K}}=-140, -150$ MeV respectively. It is seen that the onset of  (anti)kaon  condensation softens the EoS, which is marked by a change in the slope of EoSs beyond the condensation threshold. Furthermore, the softening is more pronounced in the case of NY$\Delta$ composition, which reverses the high-density behavior seen in the left panel: the EoS
with NY$\Delta$ composition is now the softest among all considered cases. It is further seen that the
higher the value of $U_{\bar{K}}$ the more pronounced is the softening of the EoSs. 

\begin{figure*} [t!]
  \begin{center}
\includegraphics[width=14.5cm,keepaspectratio ]{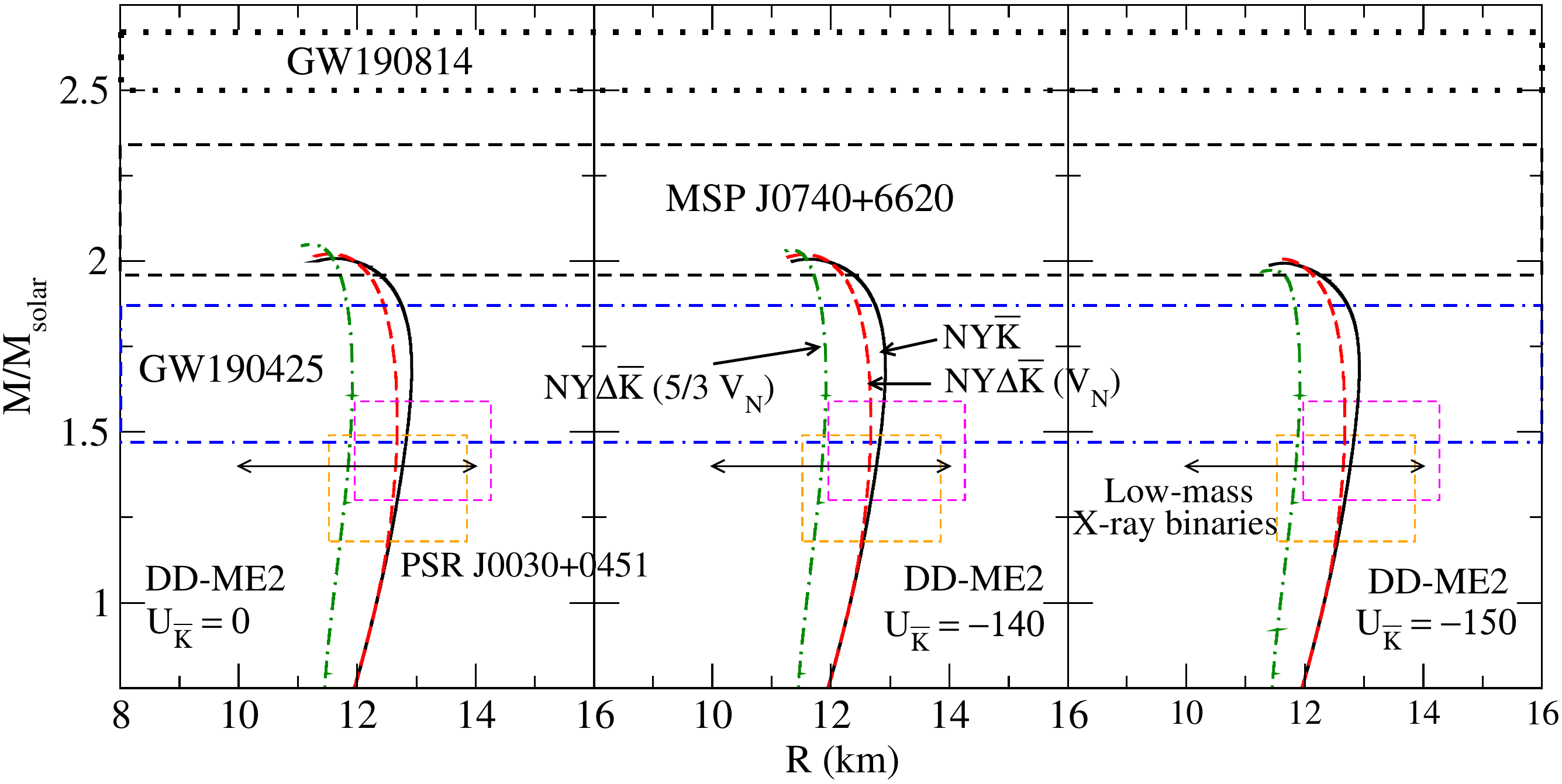}
\caption{The mass-radius relationships for EoS shown in Fig.~\ref{fig-1} for  
 NY matter (solid lines), NY$\Delta$ matter
 with $\Delta$-potential $V_\Delta=V_N$ (dashed lines) and $V_\Delta=5/3~V_N$  (dash-dotted lines). The three panels correspond to different values of (anti)kaon potential: $U_{\bar{K}}=0$, i.e., no (anti)kaon condensation,
 (left panel), $U_{\bar{K}}=-140$
  (middle panel), and $U_{\bar{K}}=-150$ MeV (right panel). 
  The astrophysical constraints from GW190425 \cite{2020ApJ...892L...3A}, GW190814 \cite{2020ApJ...896L..44A}, MSP J$0740+6620$ \cite{2020NatAs...4...72C}, PSR J$0030+0451$ \cite{2019ApJ...887L..24M,2019ApJ...887L..21R}, low-mass X-ray binaries \cite{2018MNRAS.476..421S} are shown by dot-double dashed, dotted, long-, short-dashed boxes and horizontal solid line respectively.}
\label{fig-3}
\end{center}
\end{figure*}

The mass-radius ($M$-$R$) relations corresponding to the EoSs in Fig.~\ref{fig-1} were obtained by solving the Tolman-Oppenheimer-Volkoff (TOV) equations for static non-rotating spherical stars \cite{1996cost.book.....G} and are shown in Fig.~\ref{fig-3}. For the crust region, the BPS EoS ~\cite{1971ApJ...170..299B} is used. 
The inclusion of additional exotic degrees of freedom reduces the maximum mass of NSs in comparison to nucleonic matter from $2.5~M_{\odot}$ to $\sim 2~M_{\odot}$. The compactness is also observed to be enhanced due to the appearance of $\Delta^-$-resonance at lower densities.
The parameter values of the maximum mass stars are provided in a tabulated form in Table~\ref{tab:7}. From Tables~\ref{tab:6} and \ref{tab:7} it can be inferred that $K^-$ meson appears in all the EoS models with $U_{\bar{K}}=-140, -150$ MeV. But $\bar{K}^0$ meson does not appear in the hypernuclear star with $U_{\bar{K}}=-140$ MeV and $\Delta$-baryon admixed hypernuclear star with $V_{\Delta}=V_N$ and $U_{\bar{K}}=-140$ MeV.  Consistent with the (anti)kaon softening of the EoS seen in Fig.~\ref{fig-1} the maximum masses of the stars with NY$\Delta$ composition and (anti)kaon condensation lie below those without $\Delta$ resonances, which is the reverse of what is observed when (anti)kaon condensation is absent.
\begin{table*} [t!]
\centering
\caption{Properties of maximum mass stars for various compositions and values of (anti)kaon potential $U_{\bar{K}}(n_0)$. For each composition/potential value the enteries include: maximum mass (in units of $M_{\odot}$) the radius (in units of km), and central number density (in units of $n_0$).}
\begin{tabular}{c|ccc|ccc|ccc}
\hline \hline
Configuration & \multicolumn{3}{c|}{NY$\bar{K}$} & \multicolumn{6}{|c}{NY$\Delta \bar{K}$} \\
 \cline{1-10}
 & & & & \multicolumn{3}{c|}{$V_{\Delta}=V_N$} & \multicolumn{3}{c}{$V_{\Delta}= 5/3~V_N$} \\
 \cline{5-10}
$U_{\bar{K}}$ (MeV) & $M_{max}$($M_{\odot}$) & $R$(km) & $n_{c}$($n_0$) & $M_{max}$($M_{\odot}$) & $R$(km) & $n_{c}$($n_0$) & $M_{max}$($M_{\odot}$) & $R$(km) & $n_{c}$($n_0$) \\
\hline
$0$ & $2.008$ & $11.651$ & $6.107$ & $2.021$ & $11.565$ & $6.160$ & $2.049$ & $11.226$ & $6.349$ \\
$-140$ & $2.005$ & $11.652$ & $6.096$ & $2.019$ & $11.566$ & $6.151$ & $2.032$ & $11.343$ & $ 6.214$ \\
$-150$ & $1.994$ & $11.664$ & $6.13$ & $2.006$ & $11.61$ & $6.143$ & $1.973$ & $11.448$ & $6.028$ \\
\hline
\end{tabular}
\label{tab:7}
\end{table*}

From the analysis above, we conclude that compact stars containing (anti)kaons are consistent with the astrophysical constraints set by the observations of massive pulsars, the NICER measurements of parameters of PSR J$0030+0451$, the low-mass X-ray binaries in a globular cluster, and the gravitational wave event GW190425, see Sec.~\ref{sec:intro}.  Although we do not provide here the deformabilities of our models, from the values of the radii obtained it is clear that our models are also consistent with the GW170817 event. Finally, our models are inconsistent with the interpretation of the light companion of the GW190814 binary as a compact star.  Including the rotation even at its maximal mass-shedding limit will not be sufficient to produce a $\sim 2.5M_{\odot}$ mass compact star, see Refs.~\cite{2020PhRvD.102d1301S,LI2020135812}.
\begin{figure} [h!]
  \begin{center}
\includegraphics[width=8.5cm,keepaspectratio ]{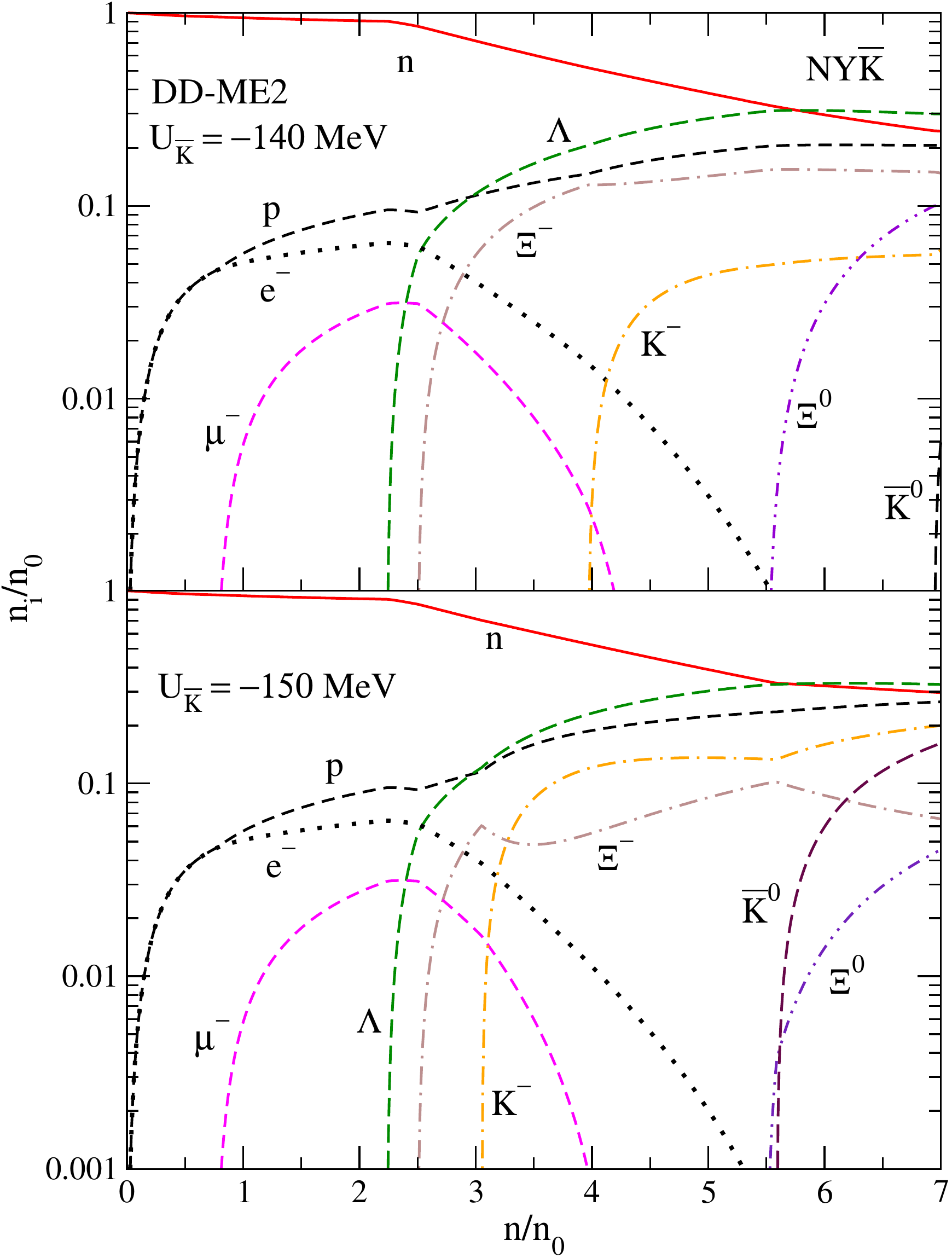}
\caption{Particle abundances $n_i$ (in units of $n_0$) as a function of normalized baryon number density in NY matter for values of  $U_{\bar{K}}-140$ MeV (top panel) and  $-150$ MeV (bottom panel). }
\label{fig-5}
\end{center}
\end{figure}

Figure~\ref{fig-5} shows the particle composition in NY matter with (anti)kaons as a function of baryon number density and for $U_{\bar{K}}=-140,-150$ MeV. At low densities, before the onset of strange particles, the charge neutrality is maintained among the protons, electrons and muons. At somewhat higher density ($\ge 2n_0$) $\Lambda$ and $\Xi^{-}$ appear in the matter (because of the repulsive nature of $\Sigma$-potential in dense nuclear matter, $\Sigma$-baryons do not appear in the composition). Finally, the (anti) kaons and $\Xi^{0}$ appear
in the high-density regime ($\ge 4n_0$). Comparing the upper and lower panels of the figure, we observe that the higher $U_{\bar{K}}$ value implies a lower density threshold of the onset of (anti)kaon, as expected.
The onset of (anti)kaons also affects the population of leptons;   $K^-$ are efficient in replacing electrons and muons once they appear, thus they contribute to the extinction of leptons, which occurs at lower densities for higher values of $U_{\bar{K}}$. In the case of $U_{\bar{K}}=-150$ MeV, the $\Xi^-$ fraction is seen to be strongly affected with the appearance of $K^-$ mesons. This is expected as $K^-$ being bosons are more energetically favorable for maintaining the charge neutrality compared to fermionic $\Xi^-$. The composition in the case of $U_{\bar{K}}=-140$ MeV, does have $\bar{K}^0$ mesons ($n_u \sim 6.95~n_0$) whereas for $U_{\bar{K}}=-150$ MeV, $\bar{K}^0$ appears at onset density $n_u \sim 5.59~n_0$ which leads to an additional softening of the EoS.
\begin{figure} [h!]
  \begin{center}
\includegraphics[width=8.5cm,keepaspectratio ]{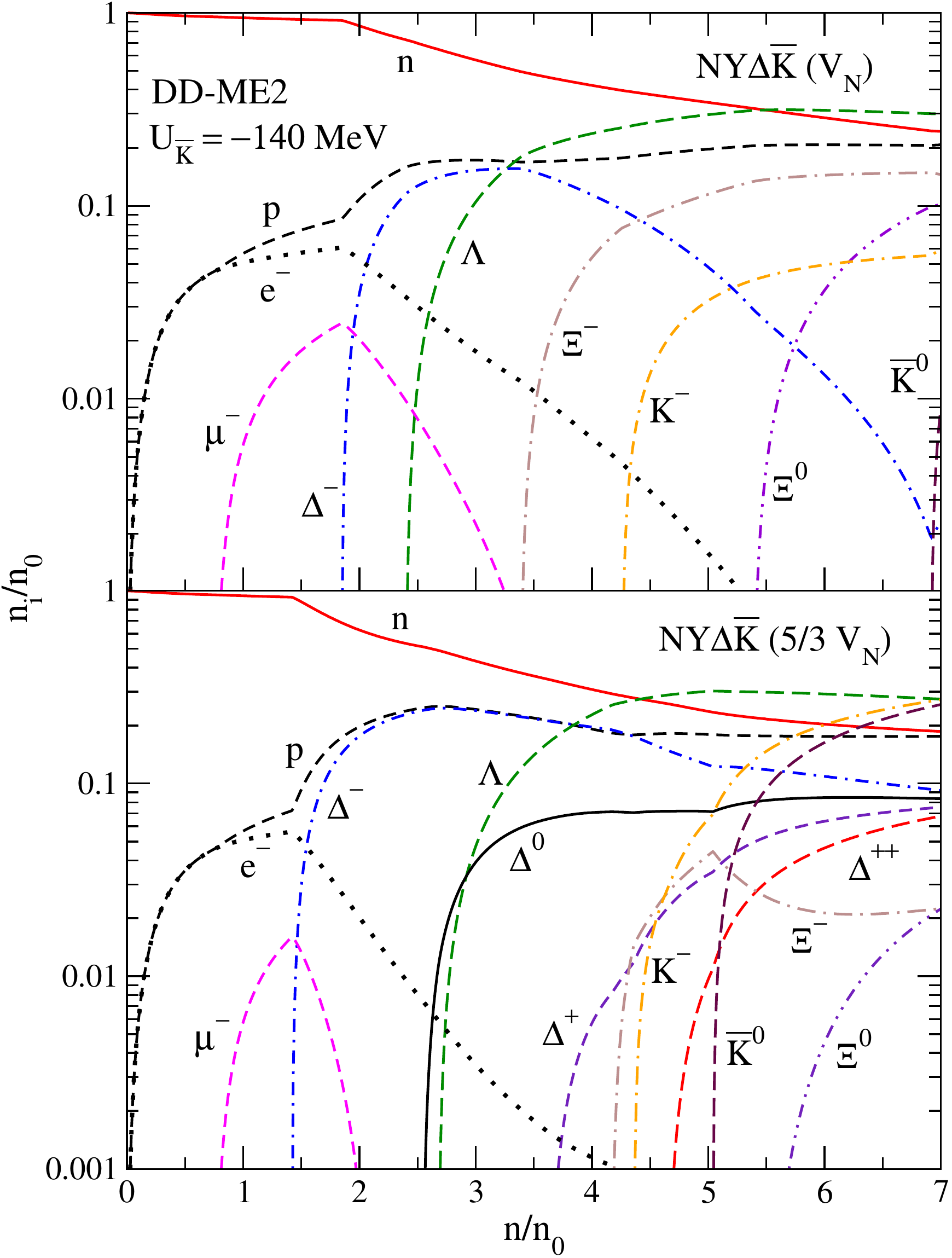}
\caption{Same as Fig.~\ref{fig-5} but for NY$\Delta$ matter
   for $V_{\Delta} = V_N$ (top panel) and $V_{\Delta} = 5/3~V_N$
  (bottom panel) and fixed value of $U_{\bar{K}}=-140$ MeV.}
\label{fig-6}
\end{center}
\end{figure}
Figure~\ref{fig-6}, which is analogous to Fig.~\ref{fig-5}, shows the particle population in $NY\Delta$-matter as a function of baryon number density for $U_{\bar{K}}=-140$ MeV. It is observed that for $V_{\Delta}=~V_N$ only $\Delta^-$ resonance appears, whereas for $V_{\Delta}=5/3~V_N$ the onset of the entire quartet of $\Delta$-resonances is possible.  It seen that in general the $\Delta$-resonances effectively shift the threshold densities of hyperons to higher densities, thus diminishing their role.  This concerns both the neutral $\Lambda$ as well as $\Xi^-$-hyperon. This shift is stronger for larger values of $V_{\Delta}$.  Resonances also suppress the lepton fraction by lowering the density at which they disappear in NY$\Delta$-matter, this effect being magnified for larger values of $V_{\Delta}$. In the high-density regime the negative charge is provided by $\Delta^-$--$\Xi^-$--$K^-$ mixture and it is seen that the rapid increase in the $K^-$ population suppresses the $\Delta^-$-$\Xi^-$ abundances for $V_{\Delta}=5/3~V_N$, as kaons are energetically more favorable than the heavy-baryons.  Note also that the onset of $\bar{K}^0$ meson abruptly decreases the abundance of $\Xi^-$, as seen in the lower panel; (in the upper panel, i.e.  for $U_{\bar{K}}=-140$ MeV and $V_{\Delta}=V_N$, the $\bar{K}^0$ mesons do not appear).  There is some qualitative differences between the two cases $V_{\Delta}=1$ and $5/3~V_N$: (a) the $\Delta^-$ baryon disappears at higher matter densities for $V_{\Delta}=1$ but its abundance is almost constant in for $V_{\Delta}=5/3~V_N$; (b) the $\Lambda$ hyperon dominates over the neutron fraction at higher density for $\sim 5.5~n_0$ in case of $V_{\Delta}=V_N$ compared to $\sim 4.5~n_0$ in case of $V_{\Delta}=5/3~V_N$.

\begin{figure} [h!]
  \begin{center}
\includegraphics[width=8.5cm,keepaspectratio ]{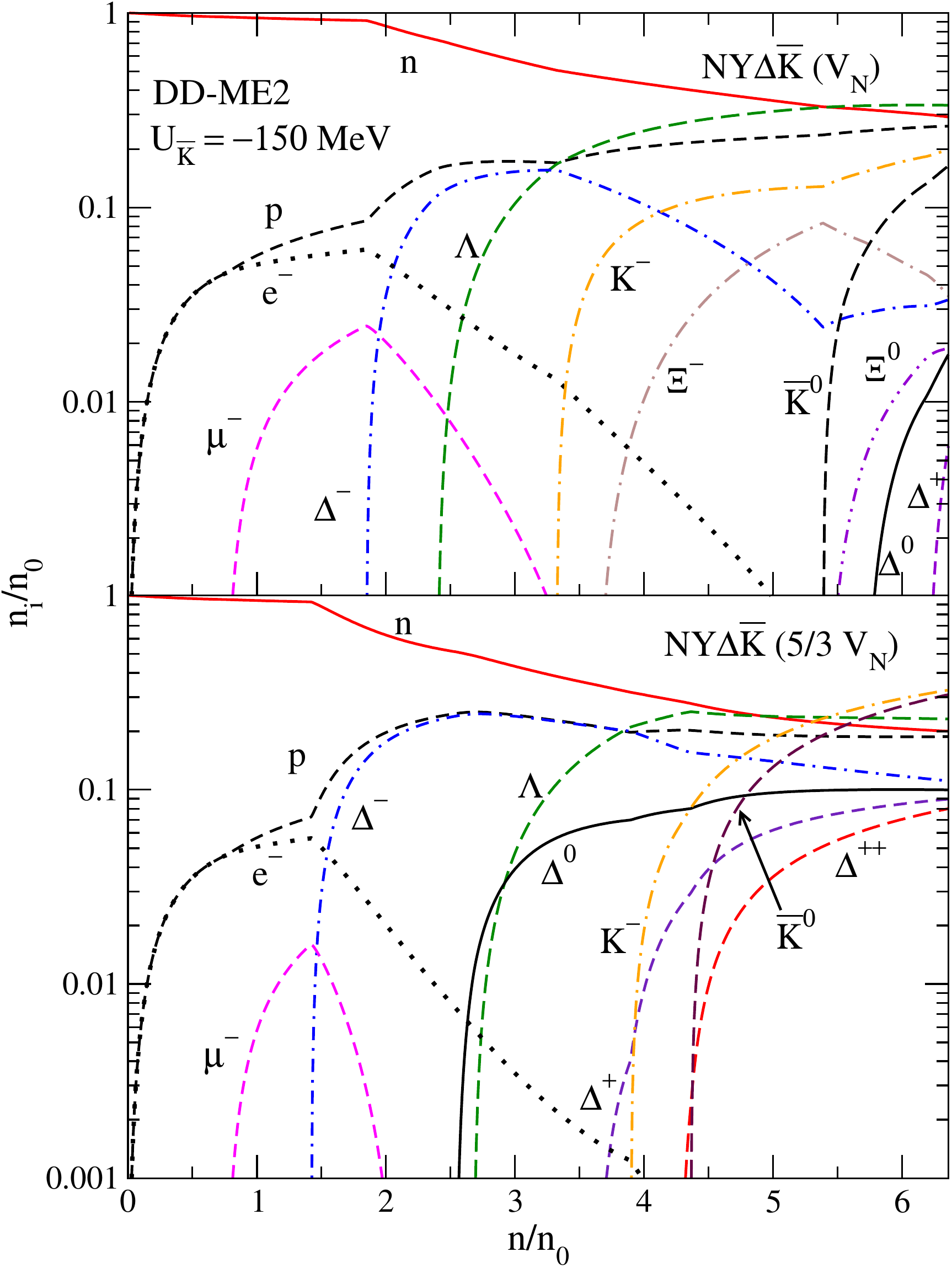}
\caption{
Same as Fig.~\ref{fig-6} but for a larger (absolute)  value of potential $U_{\bar{K}}=-150$ MeV.}
\label{fig-7}
\end{center}
\end{figure}
Figure~\ref{fig-7} shows the same as in Fig.~\ref{fig-6} but for $U_{\bar{K}}=-150$ MeV.
The particle fractions show identical trends as in Fig.~\ref{fig-6} until the appearance of (anti)kaons.
The larger potential favors earlier onset of (anti)kaons in matter; for example, the $K^-$ sets in before
the $\Xi^-$ and it is now the dominant negatively charged component shortly after the density
increases beyond the onset value.
The effect of the onset of $\bar{K}^0$ on the $\Xi^-$ and $\Delta^-$, which is mediated via changes in the abundances of $K^-$, is seen clearly again.  As before, for a large value of $V_{\Delta}=5/3~V_N$, all the members of the quartet of $\Delta$-resonances are present in the matter composition. Another notable fact is the complete extinction of $\Xi^{-,0}$ baryons, which is consistent with the trends seen in Figs.~\ref{fig-5} and \ref{fig-6}.  Interestingly, in the case $V_{\Delta}=5/3~V_N$ the (anti)kaons abundances are the largest among all particles in the high-density regime, which leads also to the softening of the EoS observed above.
\begin{figure} [h!]
  \begin{center}
\includegraphics[width=8.5cm,keepaspectratio ]{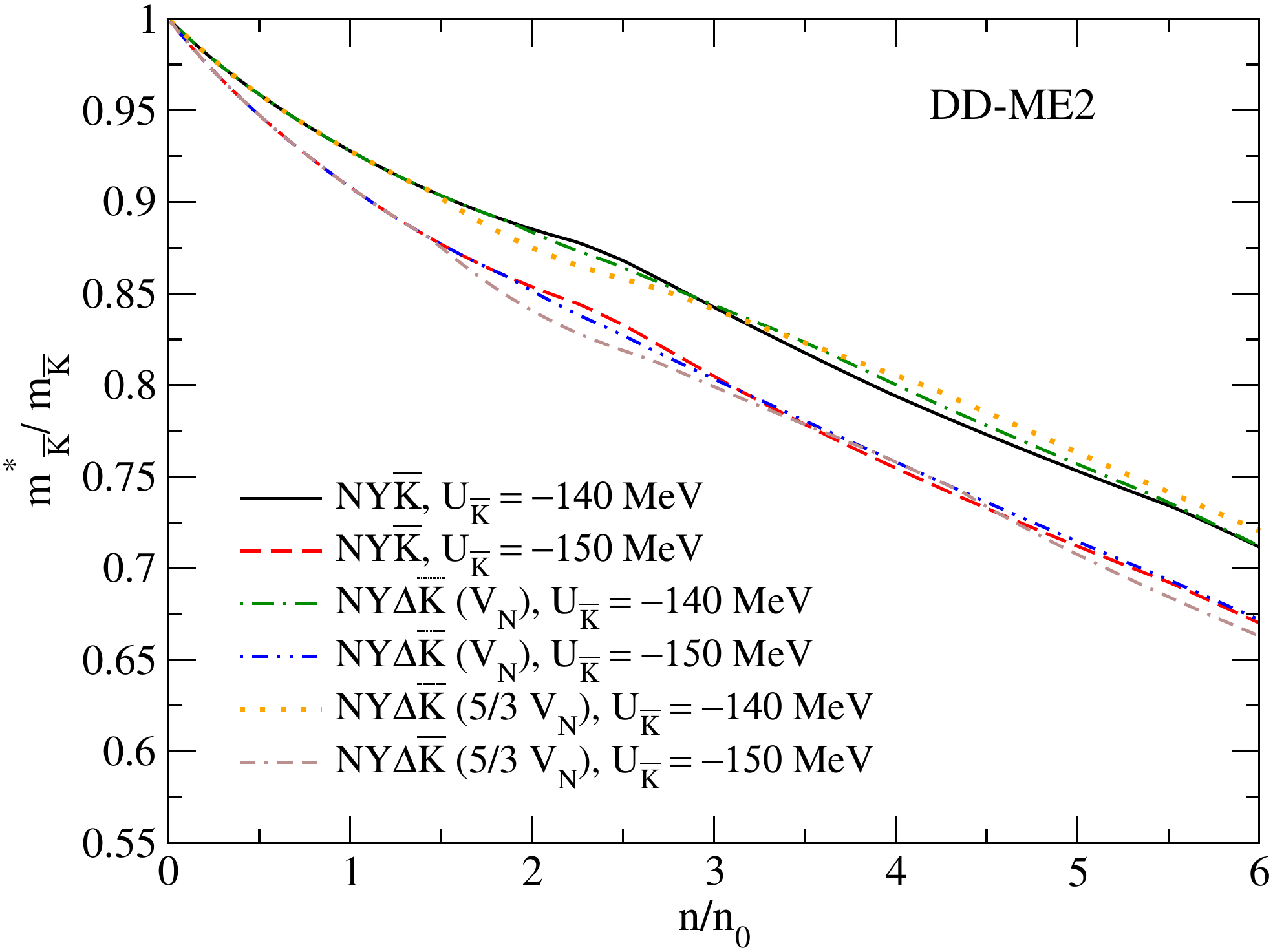}
\caption{Effective (anti)kaon mass (in units of its bare mass, $m_{\bar{K}}$) as a function of baryon number
  density for NY and NY$\Delta$ matter compositions and two values of (anti)kaon potential depth.  }
\label{fig-8}
\end{center}
\end{figure}
Figure~\ref{fig-8} shows the (anti)kaon effective mass as a function of normalized baryon number density for various strengths of $U_{\bar{K}}$ with different matter compositions. The effective mass of (anti)kaons tends to decrease rather steeply in case of higher strengths of $U_{\bar{K}}$. It is observed that in the low-density regime, the (anti)kaon effective mass decreases relatively quickly in the case of $\Delta$-resonances admixed matter compared to that with the only hyperonic matter. The reason is the larger scalar potential values arising from the onset of additional non-strange baryons at lower densities. And at higher densities, the (anti)kaon effective mass values are observed to be larger in the former case than the latter one. This may be attributed to the delayed onset of hyperons because of the $\Delta$-resonances appearance.

\begin{figure} [h!]
  \begin{center}
\includegraphics[width=8.5cm,keepaspectratio ]{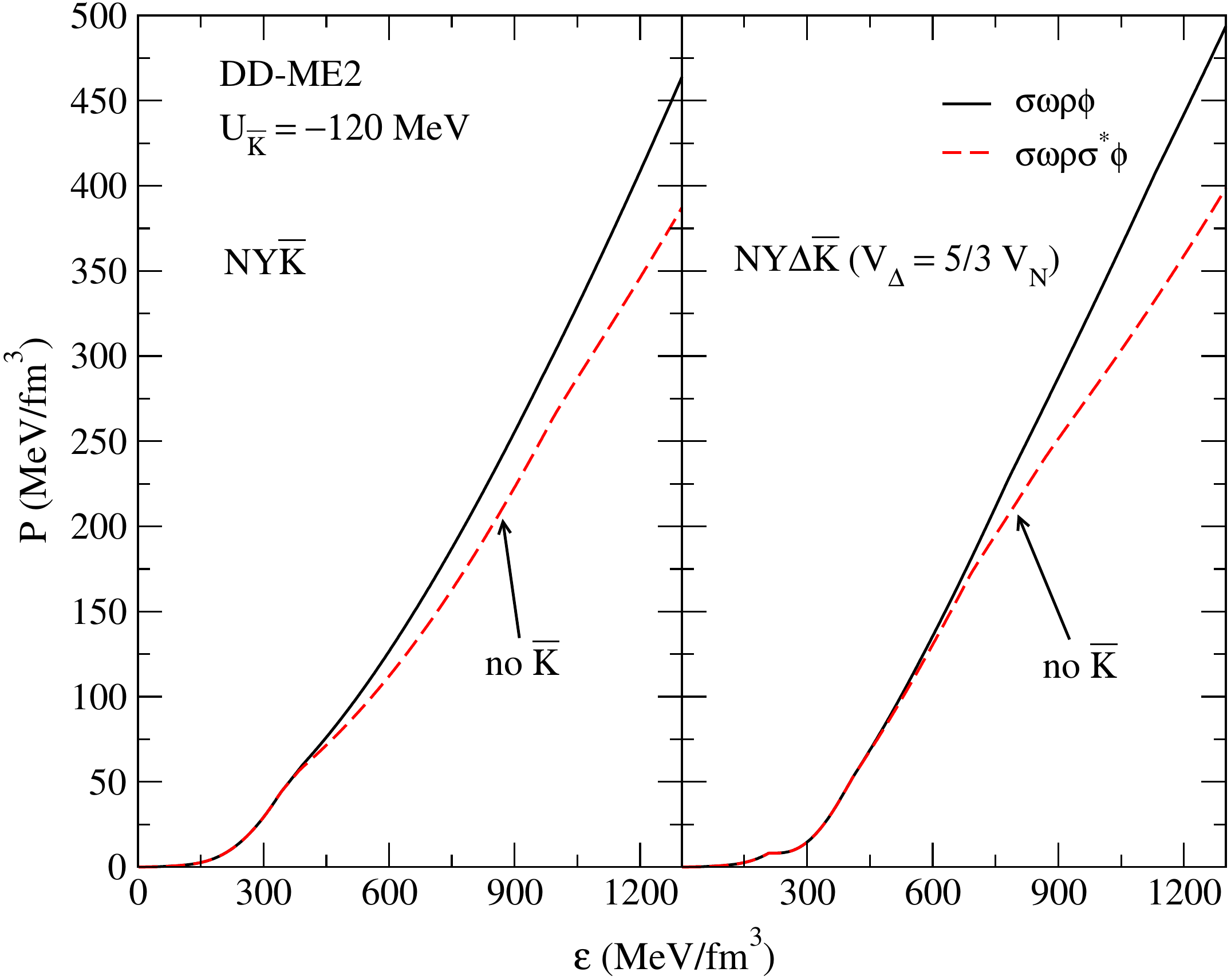}
\caption{The EoS of NY matter (left panel) and NY$\Delta$ matter (right panel) with (anti)kaon potential
  $U_{\bar{K}}=-120$ MeV including $\sigma^*$ meson (dashed lines) and without it (solid lines).
  The $\Delta$-potential value is fixed at $5/3V_{N}$.
}
\label{fig-9}
\end{center}
\end{figure}
The matter pressure as a function of energy density for different matter compositions with and without $\sigma^*$ meson for the hyperon-hyperon interactions is shown in Fig.~\ref{fig-9}. Being a scalar, $\sigma^*$ meson makes the EoS softer as is evident from the figure. It is observed that incorporating $\sigma^*$ meson rules out the possibility of (anti)kaon phase transition with $U_{\bar{K}}=-120$ MeV. This is because this scalar meson further reduces the effective mass of (anti)kaons halting their onset in the matter. The phase transition from the purely hadronic to (anti)kaon condensed phase is second-order. 
\begin{figure} [h!]
  \begin{center}
\includegraphics[width=8.5cm,keepaspectratio ]{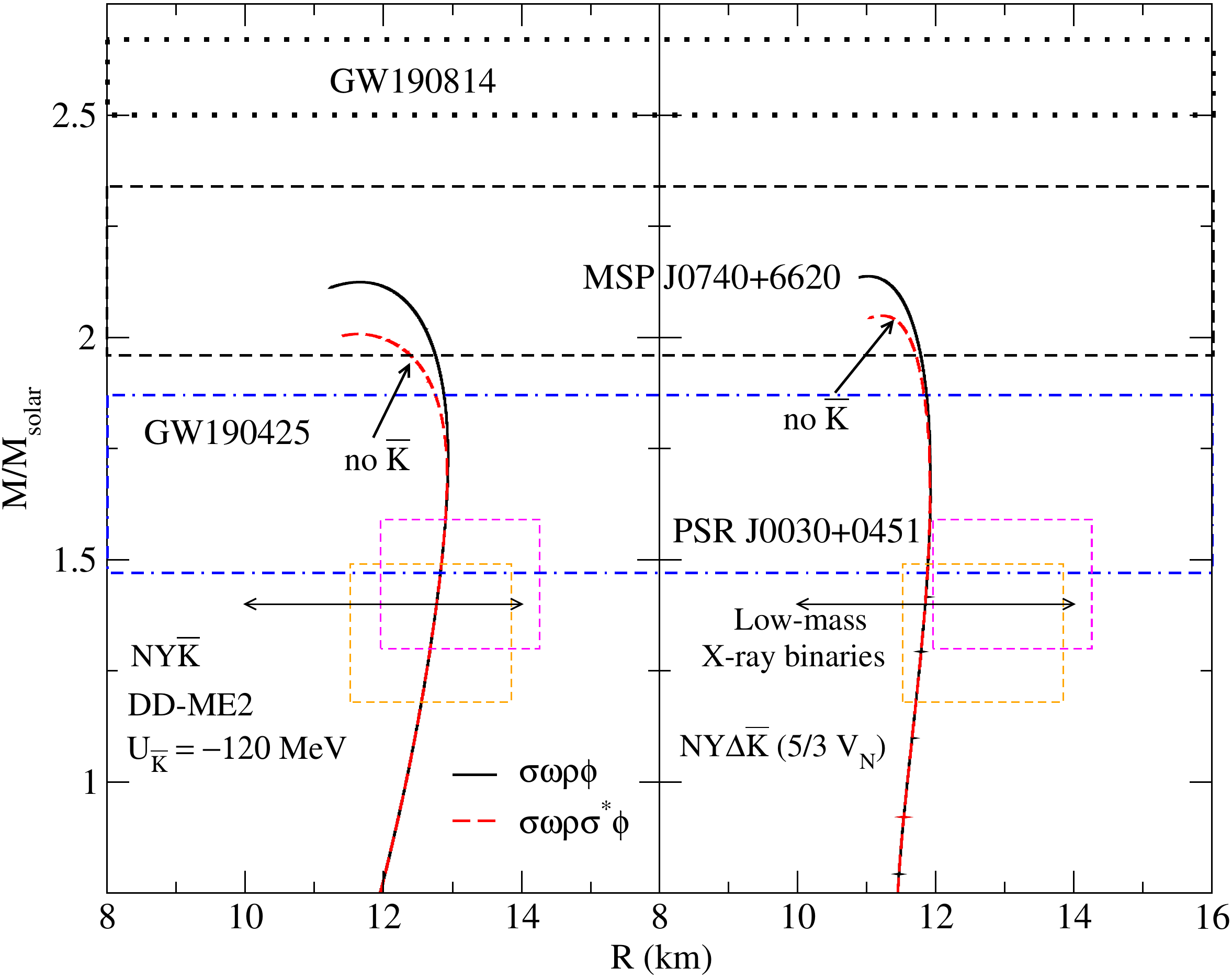}
\caption{The $M$-$R$ relations corresponding to the EoSs in Fig.~\ref{fig-9} are shown for NY matter (left panel) and NY$\Delta$ matter (right panel) with (anti)kaon potential $U_{\bar{K}}=-120$ MeV including $\sigma^*$ meson (dashed lines) and without it (solid lines).  The $\Delta$-potential value is fixed at $5/3V_{N}$.  The astrophysical observables (constraints) are similar as in Fig.~\ref{fig-3}.}
\label{fig-10}
\end{center}
\end{figure}
The results of mass-radius ($M$-$R$) relationship obtained by solving the TOV equations for non-rotating spherical stars corresponding to the EoSs in Fig.~\ref{fig-9} are presented in Fig.~\ref{fig-10}. It is observed that
in both cases of NY for NY$\Delta$ matter the inclusion of $\sigma^*$ meson leads to lower  maximum mass. It is also seen that the addition of $\Delta$'s reduces the radius of the of the stars and mildly increases the maximum, which consistent with the findings without (anti)kaon condensation. 
Table~\ref{tab:8} provides the stellear maximum masses, radii and corresponding central densities evaluated from the EoSs in Fig.~\ref{fig-9} with $U_{\bar{K}}=-120$ MeV.
\begin{table} [h!]  \centering \caption{ Properties of maximum mass stars for various compositions, $U_{\bar{K}}=-120$ MeV, $V_{\Delta}= 5/3~V_N$ in the cases with $\sigma^*$ meson and without.  In both cases we list the maximum mass (in units of $M_{\odot}$) the radius (in units of km), and central number density (in units of $n_0$).}  \begin{tabular}{c|ccc|ccc}
  \hline \hline
  Config. & \multicolumn{3}{c}{NY$\bar{K}$} & \multicolumn{3}{|c}{NY$\Delta \bar{K}$ ($V_{\Delta}= 5/3~V_N$)} \\
 \cline{1-7}
 & $M_{max}$ & $R$ & $n_{c}$ & $M_{max}$ & $R$ & $n_{c}$ \\
 & ($M_{\odot}$) & (km) & ($n_0$) & ($M_{\odot}$) & (km) & ($n_0$) \\
\hline
$\sigma \omega \rho \phi$ & $2.124$ & $11.673$ & $5.973$ & $2.137$ & $11.023$ & $6.538$ \\
$\sigma \omega \rho \sigma^* \phi$ & $2.008$ & $11.651$ & $6.107$ & $2.049$ & $11.226$ & $6.349$ \\
\hline
\end{tabular}
\label{tab:8}
\end{table}

\begin{figure} [h!]
  \begin{center}
\includegraphics[width=8.5cm,keepaspectratio ]{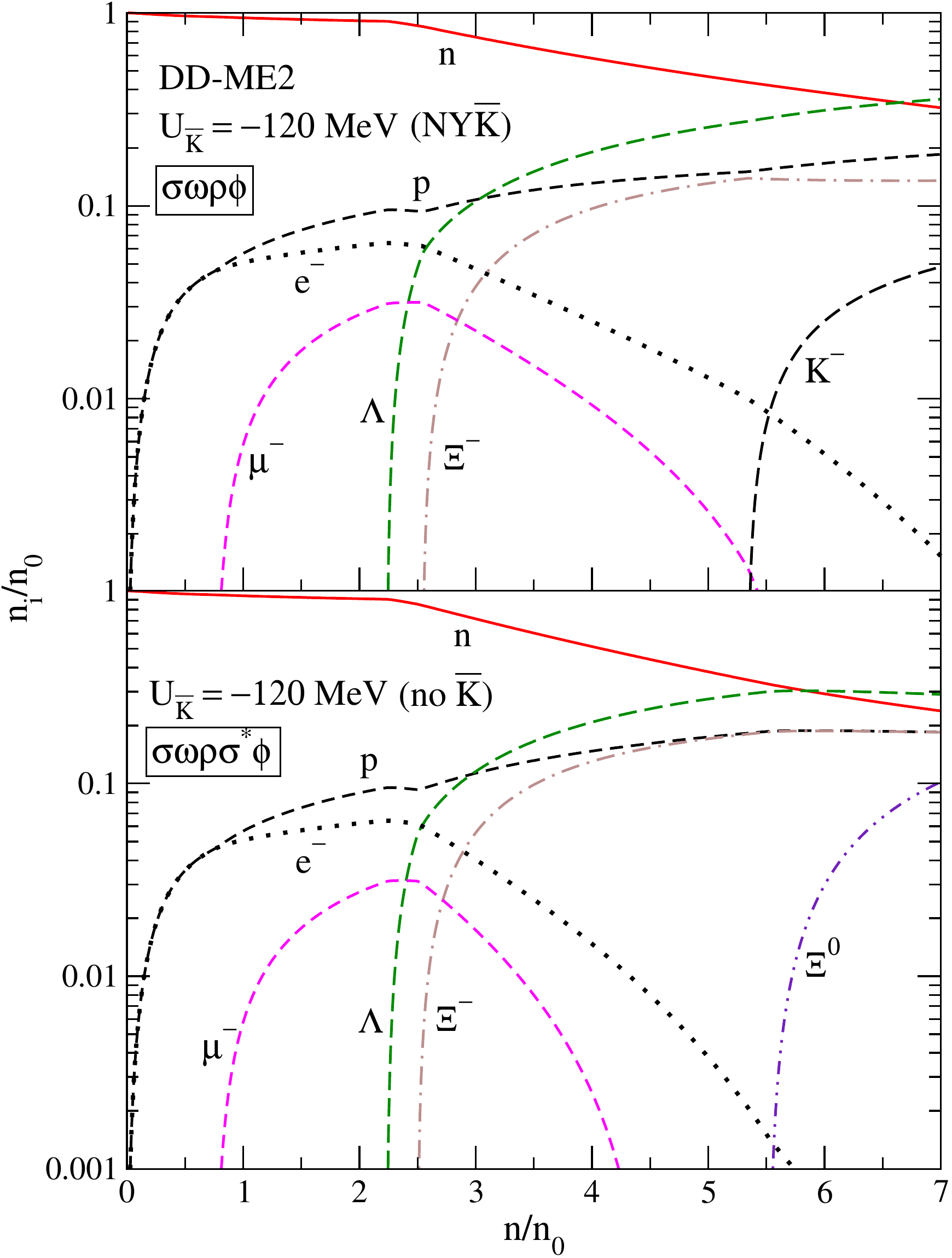}
\caption{
  Particle abundances $n_i$ (in units of $n_0$) as a function of normalized baryon number density in NY matter for value of  $U_{\bar{K}}= -120$ in the case of $\sigma \omega \rho \phi$ exchange (top panel) and
  $\sigma \omega \rho \sigma^* \phi$ (bottom panel). (Anti)kaons are absent in the second case. 
}
\label{fig-11}
\end{center}
\end{figure}
Figure~\ref{fig-11} shows the particle abundances in case of hypernuclear matter with $U_{\bar{K}}=-120$ MeV with and without $\sigma^*$ meson.  The main qualitative difference is that $K^-$ appears for $n \ge 5.4~n_0$ in the first case and it does not appear up to $n\sim 7~n_0$ in the second case. Consequently, the charge neutrality is maintained between $e-\Xi^-+K^-$ and protons in the first case and only $e-\Xi^-$ and protons in the second case. Given by more than one order of magnitude smaller abundance of electrons, the abundances of $\Xi^-$ and protons almost coincide in the second case.  Another feature seen in Fig.~\ref{fig-11} is that the electron and muon populations disappear faster with increasing density in the case where the $\sigma^*$ meson is included.

\begin{figure} [h!]
  \begin{center}
\includegraphics[width=8.5cm,keepaspectratio ]{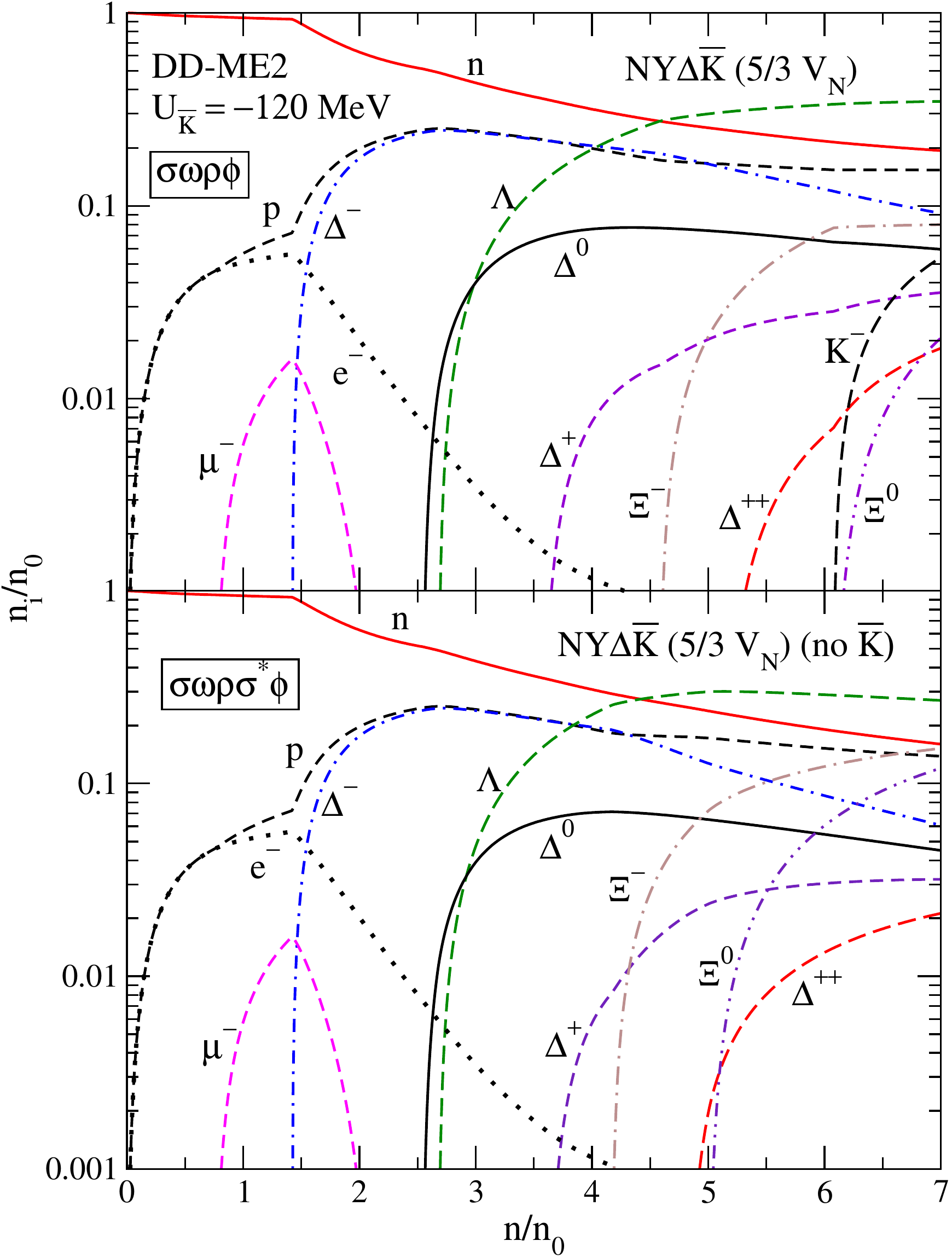}
\caption{Same as Fig.~\ref{fig-11} but for NY$\Delta$ matter with $V_{\Delta}=5/3~V_N$.}
\label{fig-12}
\end{center}
\end{figure}
Figure~\ref{fig-12}, which is similar to Fig.~\ref{fig-11}, shows the composition of particles in NY$\Delta$ matter and for $U_{\bar{K}}=-120$ MeV. In this case also, (anti)kaons are observed to appear only in the EoS where $\sigma^*$ meson is excluded.  It is seen, that the main difference between the two cases is that
$\sigma^*$ driven interactions prefer lower threshold density of $\Xi^0$ and their larger fraction, which
effectively leads to an exclusions of (anti)kaons in the density range considered. Unlike the case with only hypernuclear matter, in this case the lepton fractions are unaffected by the exclusion or inclusion of $\sigma^*$ meson, because of the negative charge is supplied by the $\Delta^-$-resonance. 

\section{Summary and Conclusions}
\label{sec:conclusions}

In this work, we discussed the second-order phase transition to Bose-Einstein condensation of (anti)kaons in hypernuclear matter with and without an admixture of $\Delta$-resonances within the framework of density-dependent CDF theory. The resulting EoS, matter composition, and the structure of the associated static, spherically symmetrical star models were presented.  The strong interactions viz. baryon-baryon and (anti)kaon-baryon are handled on the same footing. The mediators considered in this work are $\sigma$, $\omega$, $\rho$ for the non-strange baryons and two strange particle interaction mediating mesons- $\sigma^*$, $\phi$. The $K^-$ optical potentials ($-120\leq U_{\bar{K}} \leq -150$ MeV) at nuclear saturation density are considered in a range which fulfills the observational compact star maximum mass constraint ($\sim 2M_{\odot}$).

We find that the (anti)kaon condensates cannot appear in the hypernuclear matter, within our parametrization, if $U_{\bar{K}} \leq -130$ MeV. $\bar{K}^0$ condensation is absent in maximum mass compact stars with $U_{\bar{K}}=-140$ MeV. The inclusion of hyperons into the matter composition shifts the onset of (anti)kaons to higher density regimes in comparison to the case without hyperons, i.e. only nuclear matter, c.f. to Ref.~\cite{PhysRevD.102.123007}. For higher $U_{\bar{K}}$ values, the appearance of both the (anti)kaons becomes possible
in the maximum mass models. The $K^-$ meson fraction is seen to dominate over the $\Xi^-$ baryon for high $U_{\bar{K}}$ strengths. This can be attributed to the fact that
the $K^-$ particle being bosons is more favored over the fermionic $\Xi^-$-particles.

Next, in the case of $\Delta$ baryon admixed hypernuclear matter, the onset of (anti)kaons is shifted to even higher densities compared to only hyperonic matter. (Anti)kaon condensation is absent with $U_{\bar{K}} \leq -120$ MeV. The condensed phase is observed to appear in matter with $U_{\bar{K}}=-130$ MeV and $V_{\Delta}=V_N$. However, $\bar{K}^0$ condensation is absent for this particular $U_{\bar{K}}$ strength. Larger values of $\Delta$-potentials $V_{\Delta}$ imply that the entire $\Delta$-resonances quartet is present in matter.  It is also observed that in a particular matter composition ($U_{\bar{K}}=-150$ MeV, $V_{\Delta}=V_N$), the onset of $K^-$ occurs even before that of $\Xi^-$ particles. Moreover, for higher strengths of $U_{\bar{K}}$ and $V_{\Delta}$, the $\Delta$-baryons and (anti)kaons take over the $\Xi^{-,0}$ particles leading to their complete suppression in the matter. Lepton populations are suppressed with increasing density more quickly in case of higher strengths of $V_{\Delta}$.  We find that the effective mass of (anti)kaons is weakly dependent on the composition of matter and decreases almost linearly in the relevant density range $2\le n/n_0\le 6$, which reflects the density dependence of the scalar potential.

The influence of the strange scalar interaction mediating meson $\sigma^*$ on the composition and EoS are twofold: firstly, including the $\sigma^*$ meson softens the EoS significantly leading to lower maximum masses of compact stars. Secondly, exclusion of $\sigma^*$ meson allows for (anti)kaon $K^-$ to appear for weakly attractive potential strength $U_{\bar{K}}\sim -120$ MeV in both the hyperonic as well as $\Delta$ admix hypernuclear matter.

As indicated in the discussion (Sec.~\ref{sec:results}) the present model with a suitable choice of parameters characterizing the (anti)kaon condensate is consistent with the currently available
astrophysical constraints listed in Sec.~\ref{sec:intro}. The present model can, therefore, be used to model
physical processes in (anti)kaon condensate featuring $\Delta$-admixed hypernuclear star.
Examples include cooling processes, bulk viscosity, thermal conductivity, to list a few.

\begin{acknowledgements}
The authors thank the anonymous referee for the constructive comments which have bestowed to enhance the quality of the manuscript notably.
VBT and MS acknowledge the financial support from the Science and Engineering Research Board, Department of Science and Technology, Government of India through Project No. EMR/2016/006577 and Ministry of Education, Government of India. They are also thankful to Sarmistha Banik and Debades Bandyopadhyay for vital and fruitful discussions. M.S. also thanks Alexander von Humboldt Foundation for the support of a visit to Goethe University, Frankfurt am Main. J. J. Li acknowledges the A. von Humboldt Foundation for support in the initial stages of this
  work.  A. S. acknowledges the support by the Deutsche Forschungsgemeinschaft (Grant No. SE 1836/5-1)
  and  the European COST Action CA16214 PHAROS ``The multi-messenger physics and astrophysics of
neutron stars''. 
\end{acknowledgements}

\bibliography{Kaons,Kaons2,GW_ref}
\end{document}